# Generalization of a Grover based secret sharing protocol and interception attack analysis

H. Tonchev, R. Bahtev

Institute for Nuclear Research and Nuclear Energy, Bulgarian Academy of Sciences

72 Tzarigradsko Chauss´ee, Sofia, Bulgaria

E-mail: hri100t@abv.bg

***Abstract:*** *In this work, we study interception attacks against a secret sharing protocol based on Grover's search algorithm. Unlike previous works that only give the protocol for two and three participants, we have generalized the algorithm for any number of participants. The gates in the algorithm are constructed using a generalized Householder reflection. Our main goal is to obtain the probability for an eavesdropper to break the secret depending on the true initial state and the one assumed by the eavesdropper as well as on the Householder reflection phase We also deduce analytical formula for this probability in certain cases. In cases where there are two and three participants, we give an exact analytical solution. These formulas are consistent with the numerical results.*

## 1. Introduction

Classical secret sharing protocols (CSSP) [1] are protocol for distributing a secret among a group of people in a way that prevents any of them from abusing it. This is done by dividing the information into parts in such a way that none of the participants has intelligible information about the secret; only when a sufficient number of participants combine their shares, can the secret be recovered.

There are many different CSSPs, based on various mathematical principles. Each of those protocols, however, has its drawbacks. The historically first and most popular ones are Blakley's [2] and Shamir's [1] protocols. Blakley's protocol is based on the fact that *d-1* hyperplanes intersect in a point in a *d*-dimensional space, as well as on modular arithmetic. However, the protocol gives partial information about each participant's secret – each one knows that the secret is a point on their respective plane. Shamir's protocol is based on Lagrange interpolation polynomials and modulos arithmetic and its security relies on the fact that d points define a degree *(d-1)* polynomial. There are also various secret sharing protocols based on the Chinese remainder theorem [3][4].

These protocols have a variety of important applications, including the most obvious ones, such as keeping company secrets or bank accounts that can only be accessed by multiple employees together [1], secure electronic voting [5] in private companies or elections. They allow a group of employees in a company to sign a document only together, but not individually. Secure and reliable storage of information can be achieved by splitting the data and using secret sharing protocols [1]. Some modifications even allow secure storage [6][7] on a blockchain in an open online network. Recently, large databases have been widely used for various researches and for training neural networks for marketing and other types of research. Some of these databases contain confidential information that needs to be processed while the data remains secret to everybody involved. Secure multiparty computations [8] and federated learning [9] use secret sharing protocols to allow all participants to jointly evaluate a function or train a neural network on the input data, while this data remains secret.

In contrast to classical secret sharing protocols, whose security relies on various mathematical principles, quantum secret sharing protocols' security is relies on the laws of physics. Quantum effects like superposition, entanglement or interferences thogeter with the no-cloning theorem are used for obtaining secure cryptographic protocols. Some examples for such protocols are the ones based on quantum Fourier transform [10], quantum teleportation [11], graph states [12][13], quantum random walk [14][15] and Grover's search algorithm [16].

Grover's quantum search algorithm [17], is a black box algorithm that achieves quadratic speedup in searching in an unordered database compared to the classical search algorithms. It was first invented in 1996 [18], and is the second major quantum algorithm after Shor's factoring algorithm [19]. Together they ignite interest in quantum information by showing that a quantum computer can be used to tackle important problems faster than classical algorithms. The original Grover's search is probabilistic, but it can be made deterministic by using generalized Householder reflections [20]. These reflections can be used to efficiently decompose any operator [21]. They can also easily to be implemented experimentally in various systems including ion traps [22] and photonic quantum computers [23]. Grover's algorithm has a smaller register size than other quantum algorithms for unordered database search. Due to its simplicity, Grover's operator is used as a component in other quantum algorithms [24][25], and in quantum cryptography [16].

A secret sharing protocol based on Grover's algorithm was first introduced by Hsu [16]. Its security was improved [26] and protocol was experimentally demonstrated using nuclear magnetic resonance [27]. Recently, a modification for one distributor and 3 participants was introduced by Rathi et al. [28].

Previous works give numerical simulations about the security of the protocol and only in the cases when standard Householder reflections are used. In this paper, we give both analytical and numerical solutions about the security against interception attacks in the case of using generalized Householder reflections for two and three participants and for certain cases for an arbitrary number of participants.

This paper is organized as follows: In Section 2, the Grover's algorithm is described. Firstly, in Section 2.1, we describe its procedure and quantum circuit. In Section 2.2, we show how the required number of iterations and the probability of obtaining a solution depend on the quantum register's size. In Section 2.3 we discuss a modification of Grover's algorithm that uses generalized Householder reflections to reduce the algorithm's theoretical failure rate to zero. In Section 3 mutually unbiased bases are briefly described. Section 4 shows the scheme of the secret sharing protocol based on Grover's algorithm. The cases of two and three participants are explained in Subsection 4.1 and Subsection 4.2 respectively. Interception attacks are briefly discussed in Subsection 4.3. Our new results begin in Section 5. In the following subsections we give analytical solution in the case of two participants (Subsection 5.1) and for two variants of the protocol for the case of 3 participants. The first variant of the protocol and its security against interception attack is shown in Subsections 5.2.1 and 5.2.2 respectively. The second variant and its security against interception attacks are shown in Subsections 5.2.3. Section 6 gives a generalization of the protocol for an arbitrary number of participants. In Subsection 6.1. we show that in the case of an arbitrary number of participants, if all participants are honest, they will always find the secret (up to a phase factor) by applying Grover's algorithm to the initial state. In Subsections 6.2, and 6.3. we deduce analytical formulas for the probability of obtaining the secret when the distribution of the secret between the participants is done during the first and during the last Grover's iteration respectively. Subsections 6.4. describes the general case. Subsections 6.5 shows semi-empirical evaluation for the security of these protocols. In Section 7 the advantages and disadvantages of the secret sharing protocol with more than two participants are briefly discussed. The paper finishes with a conclusion in Section 8.

## 2. Grover's Algorithm

There are various quantum algorithms for searching in an unordered database. These include the Grover's search algorithm [17], the fixed point quantum search algorithm [29][30] and the quantum random walk search algorithm [24]. Each of these algorithms finds the searched element quadratically faster than the best known classical unordered search algorithm and each one has different advantages and disadvantages. For example, a quantum random walk search can find a node in a graph with an arbitrary topology, but it requires twice as many iterations and larger register size. Fixed-point quantum search requires a larger register size and classical measurements during the execution of the algorithm, but exceeding the necessary number of iterations does not reduce the probability of finding a solution. Grover's algorithm can only be used for searching in linear database. However, it requires a smaller quantum register size and is easy to implement experimentally [21] in various systems, including photonic [23] and ion-trap quantum computers [22]. The original algorithm is probabilistic, but can be modified to be made deterministic [20]. In the next few subsections, Grover's algorithm is briefly explained.

### 2.1. Procedure

The quantum circuit of Grover's algorithm is shown in Fig.1. The description of the quantum gates used in it is shown below.

The algorithm uses a single quantum register of dimension d for an n-qubit register $d = 2^n$. Let us denote the quantum state vectors' dimensions with subscripts e.g. $|0\rangle_d$ is d-dimensional vector. The initial state of the register in Grover's algorithm is:

$$|0\rangle_d = |0\rangle_2^{\otimes n} \quad (1)$$

The register is then put in an equal superposition of all basis states $|j\rangle_d, j = 0,1, \dots, d-1$ using a discrete Fourier transform $F_j$ or any $d$-dimensional unitary matrix in which the first column consists of elements with equal moduli.

$$|\psi\rangle_d = F_j|0\rangle_d = \frac{1}{\sqrt{d}}\sum_{j=0}^{d-1}|j\rangle_d \quad (2)$$

Next, the Grover iteration is applied multiple times to the register. The iteration consists of the following gates:

1) The oracle operator $O$. It is constructed using a function that can recognize the solution states and marks them by shifting their sign. At the beginning of the $r$-th iteration, the state of the register is:

$$|\psi_r\rangle_d = \frac{1}{\sqrt{\sum_{j=0}^{d-1}|\lambda_j\ (r)|^2}}\sum_{j=0}^{d-1}\lambda_j(r)|j\rangle_d \quad (3)$$

where $\lambda_j\ (r)$ are the coefficients in front of the $j$-th basis vector $|j\rangle_d$ at the beginning of the $r$-th iteration. Then, after the oracle is applied, the state becomes:

$$O|\psi_r\rangle_d = \frac{1}{\sqrt{\sum_{j=0}^{d-1}\left|\lambda_j\ (r)\right|^2}}\left(\left(\sum_{j=0}^{d-1}\lambda_j(r)|j\rangle_d\right) - 2\lambda_M(r)|\alpha\rangle_d\right) \tag{4}$$

where $(r)|\alpha\rangle_d$ is equal superposition of all solution states.

$O$ does a reflection on $\left|\psi_r\right\rangle_d$ against the equal superposition of all non-solution states. It can be expressed as a Householder reflection operator:

$$O = I_d - 2|\alpha\rangle_d\langle\alpha|_d \tag{5}$$

where $I_d$ is an identity matrix of dimension $d$.

2) A reflection against the initial state $\left|\psi_0\right\rangle_d$ of the algorithm.

$$U_0 = I_d - 2\left|\psi_0\right\rangle_d\left\langle\psi_0\right|_d \tag{6}$$

For the Grover iteration we have:

$$G = U_0 O. \tag{7}$$

After $G$ is applied a certain number of times a measurement is done on the register. The measurement has a high probability of returning a solution, but a non-zero chance that it does not. It should be noted that the state of the algorithm register must be measured exactly when the required number of iterations have been performed. The probability of finding a solution is a periodic function of the number of iterations. If more or fewer than the necessary amount of iterations are done, this probability decreases.

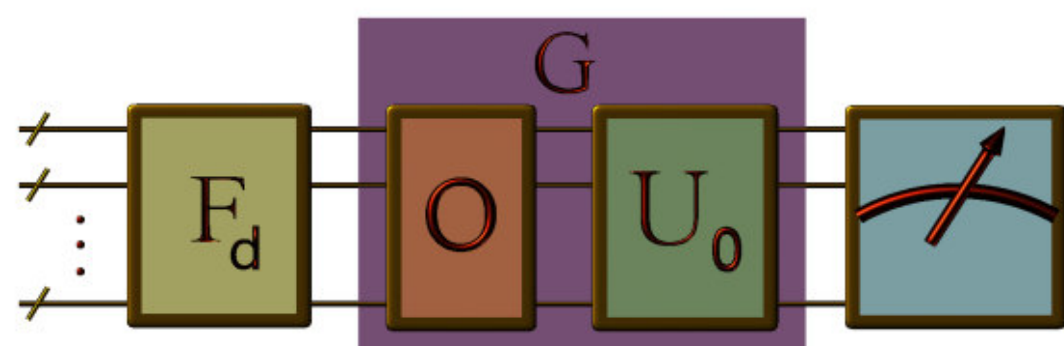

*Fig.1. Quantum circuit of Grover's search algorithm. Each of the quantum gates is shown in a different colour: Fourier transform $F_d$ operator in dark yellow, The oracle $O$ in red, the reflection operator $U_0$ in green and the measurement in teal. The iteration of the algorithm $G$ is marked with a rectangle around the $O$ and $U_0$ gates.*

In the next subsection, we deduce the number of iterations and the probability of obtaining a solution.

### 2.2. Probability of obtaining a solution and number of iterations needed

Both the operators $O$ and $U_0$ are reflections, and thus the Grover iteration $G$ is a rotation in the plane spanned by two vectors: the equal superposition of all non-solution states and the equal superposition of all solution states. In the case of only one solution, the angle by which $G$ rotates the state vector is:

$$\Theta = 2arcsin\left(\frac{1}{\sqrt{d}}\right) \tag{8}$$

For the searched element to be obtained with certainty after $k'$ iterations, the following equality must be fulfilled:

$$\frac{\pi}{2} = \left(k' + \frac{1}{2}\right)\Theta \tag{9}$$

The number of iterations is an integer, so it is equal to the smallest integer $k$ greater than $\pi/2\Theta - 1/2$. It can be shown that $k$ can be expressed as:

$$k = \left\lceil \frac{\pi}{4}\sqrt{d} \right\rceil \tag{10}$$

where $\lceil \quad \rceil$ denotes rounding up.

In general, the probability $P$ of obtaining a solution is less than one because $k = \lceil k' \rceil > k'$, so the quantum algorithm rotates the state vector more than necessary. With the increase in the register size the angle $\Theta$, and hence the value of $|\mathrm{k} - k'|$, becomes smaller, leading to a higher probability $P$:

$$P = 1 - O\left(\frac{1}{d}\right) \tag{11}$$

In the next subsection we show way the algorithm can be modified in a way that a solution is obtained with certainty.

### 2.3. Grover's algorithm with zero theoretical failure rate

One way of achieving probability $P = 1$ of finding a solution a is by replacing the operators $O$ and $U_0$ is the algorithm with generalized Householder reflection. This reduces the rotation angle of the Grover iteration $\Theta$, which depends on $\Omega$ [20]. In order to achieve a probability of finding a solution equal to one, the phases in operators $O(\Omega)$ and $U_0(\Omega)$must match [31]:

$$O(\Omega) = I_d - \left(1 - e^{i\Omega}\right)|M\rangle_d\langle M|_d \tag{12}$$

$$U_0(\Omega) = I_d - \left(1 - e^{i\Omega}\right)|\psi_0\rangle_d\langle\psi_0|_d \tag{13}$$

Thus the Generalization becomes:

$$G(\Omega) = O(\Omega)U_0(\Omega) \tag{14}$$

The maximum rotation of Grover's iteration is achieved when $\Omega = \pi$. In this case, the modified Grover rotation angle is equal to the rotation angle in the unmodified Grover rotation. If $\Omega = 0$, then there is no rotation at all. In the case of only one solution, the optimal phase needed to obtain a solution with certainty ($\Omega'_{MAX}$) can be calculated as shown in [20]:

$$\Omega'_{MAX} = 2\arcsin\left(\frac{sin\left(\frac{\pi}{4J_{\Omega'}+6}\right)}{d^{-\frac{1}{2}}}\right) \tag{15}$$

where $J$ is calculated by the formula:

$$J_{\Omega'} = \left\lfloor \frac{0.5\pi - \Theta}{2\Theta} \right\rfloor \tag{16}$$

where ⌊ ⌋ denotes rounding down t the number in the brackets.

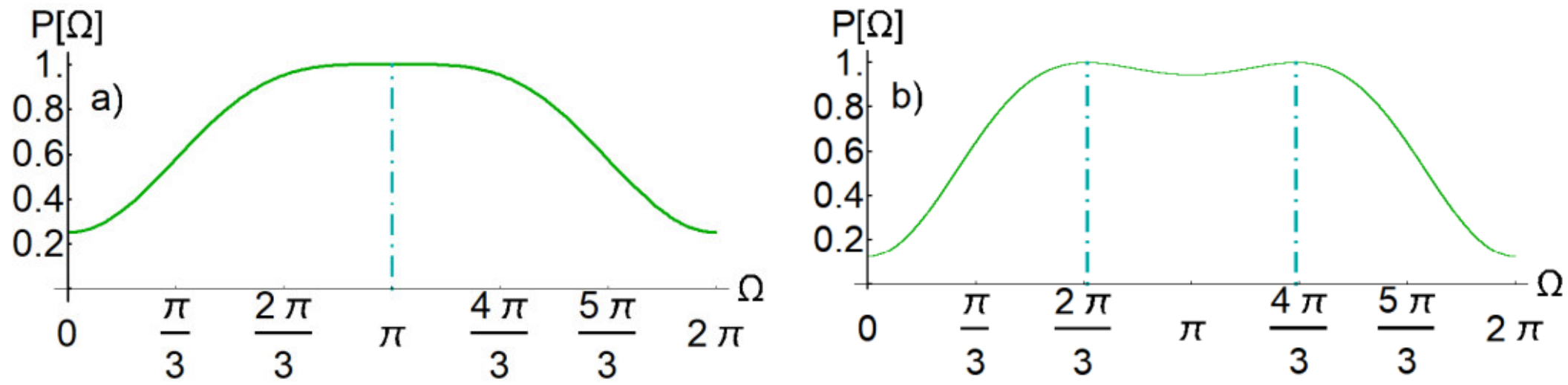

*Fig.2. Probability for Grover's search to obtain a solution as a function of the phase Ω used in both operators, for register sizes 4 (plot a) and 8 (plot b). The teal dash-dotted line marks the values of Ω for which is obtained a solution with certainty.*

*Fig.2* shows numerical simulations of the probability $P(\Omega)$ of finding a solution depending on the generalized Householder reflection phase, for register sizes 4 and 8 respectively. Note that for register size 8 there are in fact two values of Ω for which $P(\Omega) = 1$, namely $\Omega = \Omega'_{MAX}$ and $\Omega = 2\pi - \Omega'_{MAX}$ (where $\Omega'_{MAX}$ is obtained by using Eq. (15)).

*Table1* shows the values of $P(\Omega'_{MAX})$ and $P(\pi)$ for various register sizes.

| Qubits | Register size | $\Omega'_{\mathrm{MAX}}$ | $\mathrm{P}(\Omega'_{\mathrm{MAX}})$ | $\mathrm{P}(\pi)$ |
|---|---|---|---|---|
| 2 | 4 | π | 1 | 1 |
| 3 | 8 | 2.12688 | 1 | 0.945313 |
| 4 | 16 | 2.19911 | 1 | 0.961319 |
| 5 | 32 | 2.76774 | 1 | 0.999182 |
| 6 | 64 | 2.60752 | 1 | 0.996586 |

*Table1. Values of $P(\Omega'_{MAX})$ and $P(\pi)$ for various register sizes.*

### 3. Mutually unbiased bases

Two bases are mutually unbiased if measuring any basis state in one of the base in the other base does not reveal any information about this state and vice versa [32]. In a two dimensional vector space the following bases are mutually unbiased:

$$|0\rangle_2 \qquad\qquad |1\rangle_2 \tag{17}$$

$$|+\rangle_2 = (|0\rangle_2 + |1\rangle_2)/\sqrt{2} \qquad\qquad |-\rangle_2 = (|0\rangle_2 - |1\rangle_2)/\sqrt{2} \tag{18}$$

$$|+i\rangle_2 = (|0\rangle_2 + i|1\rangle_2)/\sqrt{2} \qquad\qquad |-i\rangle_2 = (|0\rangle_2 - i|1\rangle_2)\,/\sqrt{2} \tag{19}$$

It is impossible to construct fourth basis mutually unbiased to all of the above.

## 4. Grover-based secret sharing protocol for two and three participants

Grover's algorithm has been used to construct secret sharing protocols for two [16] and three participants [28]. It can only be used for encrypting classical information. For this protocol to be secure, it is required that at least one of the participants be honest.

### 4.1. Grover-based secret sharing protocol with two participants

Let Dan (D) be the distributor of the secret, who needs to share a secret integer with Alice (A) and Bob (B). However, he suspects that one of them may be dishonest. One's honesty is guaranteed as operations are performed collectively by both participants.

Initially the secret integer is converted to a binary string, which is then parsed into groups of two binary digits each. After that, based on this binary representation, each group are represented as a tensor product of qubits.

For example, let the secret number be 23 with binary representation 10111. The number of binary digits in this string is not multiple of two, so a zero is added in front of it, thus the secret becomes 010111. The latter is parsed into three groups: $M_1 = 01$, $M_2 = 01$ and $M_3 = 11$. For each group $M_j$ a two-qubit state is constructed as:

$$\begin{aligned} |M_1\rangle_4 &= |0\rangle_2 \otimes |1\rangle_2 = |1\rangle_4 \\ |M_2\rangle_4 &= |0\rangle_2 \otimes |1\rangle_2 = |1\rangle_4 \\ |M_3\rangle_4 &= |1\rangle_2 \otimes |1\rangle_2 = |3\rangle_4 \end{aligned} \tag{20}$$

and the following procedure is done:

1) Dan randomly selects the initial state of each qubit in the register ($|S_A\rangle_2$ and $|S_B\rangle_2$) among the states $|+\rangle_2$, $|-\rangle_2$, $|+i\rangle_2$ and $|-i\rangle_2$, without disclosing this initial state:

$$|S_j\rangle_4 = |S_A\rangle_2 \otimes |S_B\rangle_2 \tag{21}$$

An example of such state is $|S\rangle_4 = |+\rangle_2 \otimes |-i\rangle_2$. The number of possible initial states is 16.

2) The following operator is constructed:

$$U_{M,j} = I_4 - 2|M_j\rangle_4\langle M_j|_4 \tag{22}$$

It is then applied on the initial state to encode the message.

$$|X_j\rangle_4 = U_{M,j}|S_j\rangle_4 \tag{23}$$

The second step completes the encryption procedure. The two qubits in the encoded message $|X_j\rangle_4$ are separated without being measured. The first of them is sent to Alice and the second one to Bob. After each participant receives a qubit, they notify Dan on an open classical channel. If any of the participants has not received a qubit, the protocol is terminated, and if any participant lies about not receiving a qubit, the protocol is also terminated.

Decoy states can be prepared in the $\{|0\rangle, |1\rangle\}$ basis. This allows an honest participant to detect an eavesdropping, and to notify the distributor on the open classical channel. The distributor terminates the procedure if there is any indication that security has been compromised. If there are also decoy states prepared using the same bases as these used to encode the message, it improves the security of the protocol [26]. In this case the distributor can detect the eavesdropper at the end of the protocol procedure.

After Alice and Bob each receive a qubit and the decoy states indicate that there is no eavesdropping attempt to intercept the qubits, over the open classical channel D announces to them their respective initial states.

When both Alice and Bob know the initial states of their qubits, they can obtain the secret by acting together. They gather together their qubits and use these states to construct the operator:

$$U_{S,j} = I_4 - 2\left|S_j\right\rangle_4\left\langle S_j\right|_4 \qquad (24)$$

The Grover's algorithm with register size four is deterministic and requires only one iteration. Therefore, decoding the message is done only by applying $U_{S,j}$:

$$\left|M_j\right\rangle_4 = U_{S,j}\left|X_j\right\rangle_4 \qquad (25)$$

This completes the only iteration of Grover's algorithm. After measurement the secret $M_j$ is obtained with certainty.

The entire procedure for encoding and decoding the message is shown in *Fig.3*

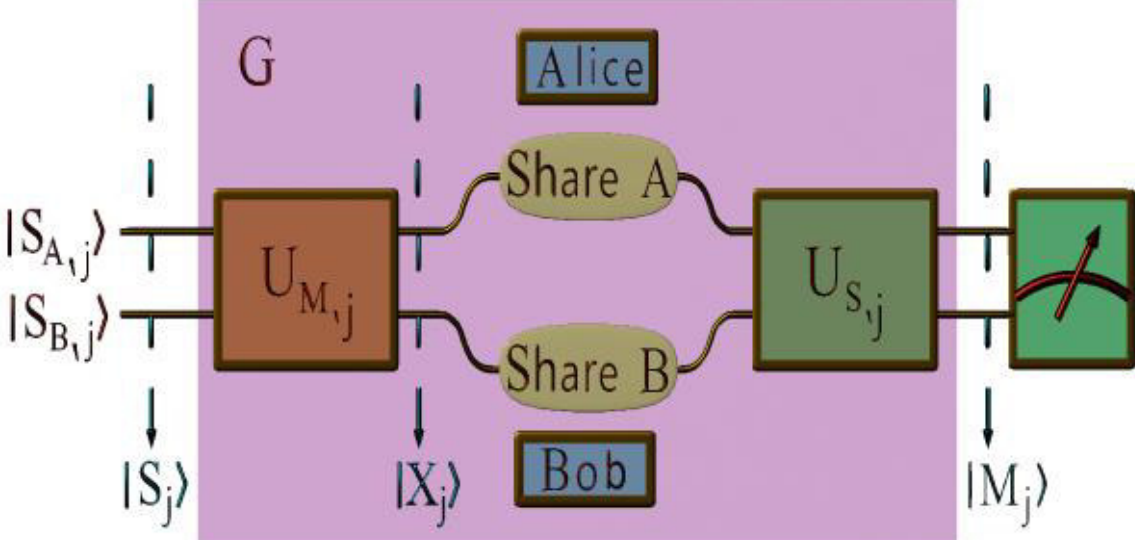

*Fig.3. Schematic of a secret sharing protocol with one distributor and two participants.*

Dan, Alice, and Bob then repeat this procedure for each pair of bits in the message.

#### 4.2. Grover-based secret sharing protocol with three participants

Suppose the distributor Dan (D) must share a secret integer with Alice (A), Bob (B) and Charlie (C). He suspects that at least one of them is dishonest. The protocol encodes the message in way that the meaningful information can only be obtained when all participants combine their shares.

For example, let the secret integer be 125, with binary representation: $M = 1111101$. The number of binary digits in this string is not a multiple of three, so two zeros are added in front of it: $001111101$. The latter string is then parsed into three group $M_j$ into three groups: $M_1 = 001$, $M_2 = 111$ and $M_3 = 101$. For each group $M_j$ a qubit state is constructed as follows:

$$|M_1\rangle_8 = |0\rangle_2 \otimes |0\rangle_2 \otimes |1\rangle_2 = |1\rangle_8$$
$$|M_2\rangle_8 = |1\rangle_2 \otimes |1\rangle_2 \otimes |1\rangle_2 = |7\rangle_8 \quad (26)$$
$$|M_3\rangle_8 = |1\rangle_2 \otimes |0\rangle_2 \otimes |1\rangle_2 = |5\rangle_8$$

The procedure for encoding the secret has the following steps:

1) D randomly selects the initial states of each qubit in the register among the states $|+\rangle_2$, $|-\rangle_2$, $|+i\rangle_2$ and $|-i\rangle_2$. He does not disclose those states.

$$|S_j\rangle_8 = |S_A\rangle_2 \otimes |S_B\rangle_2 \otimes |S_C\rangle_2 \quad (27)$$

The number of possible initial states is 64. One example for such a state is $|S\rangle_8 = |-i\rangle_2 \otimes |+\rangle_2 \otimes |-i\rangle_2$.

2) The encoding operator is constructed the same way as in the case of two participants.

$$U_{M,j} = I_8 - 2|M_j\rangle_8\langle M_j|_8 \quad (28)$$

Then $U_{M,j}$ is applied to the initial state:

$$|X_j\rangle_8 = U_{M,j}|S_j\rangle_8 \quad (29)$$

One qubit from each $|X_j\rangle_8$ is then sent to each participant. After they all confirm receiving them, D announces on an open channel the initial state of each qubit.

In the case of two participants, they simply had to complete the Grover's iteration by constructing and applying $U_{S,j}$. In order to construct the operator, the only requirement is to know the initial states of each qubit. Participants can apply these operators "anywhere", and will obtain the secret with probability 1.

To complete the procedure, the participants need to perform an additional Grover's iteration after the first one. For them to be able to do it without knowing the secret $|M_j\rangle_8$, the secret must be locked in a "padlock" from which it cannot be extracted. For the rest of the current work assume that this is the case. Participants must combine their qubits and feed the initial states into the "padlock", which then executes the rest of the protocol procedure.

The decryption of the encrypted message from the padlock can be written as:

$$|M_j\rangle_8 = U_{S,j}U_{M,j}U_{S,j}|X_j\rangle_8 \quad (30)$$

The entire encryption and decryption procedure can be depicted as shown in *Fig.4*.

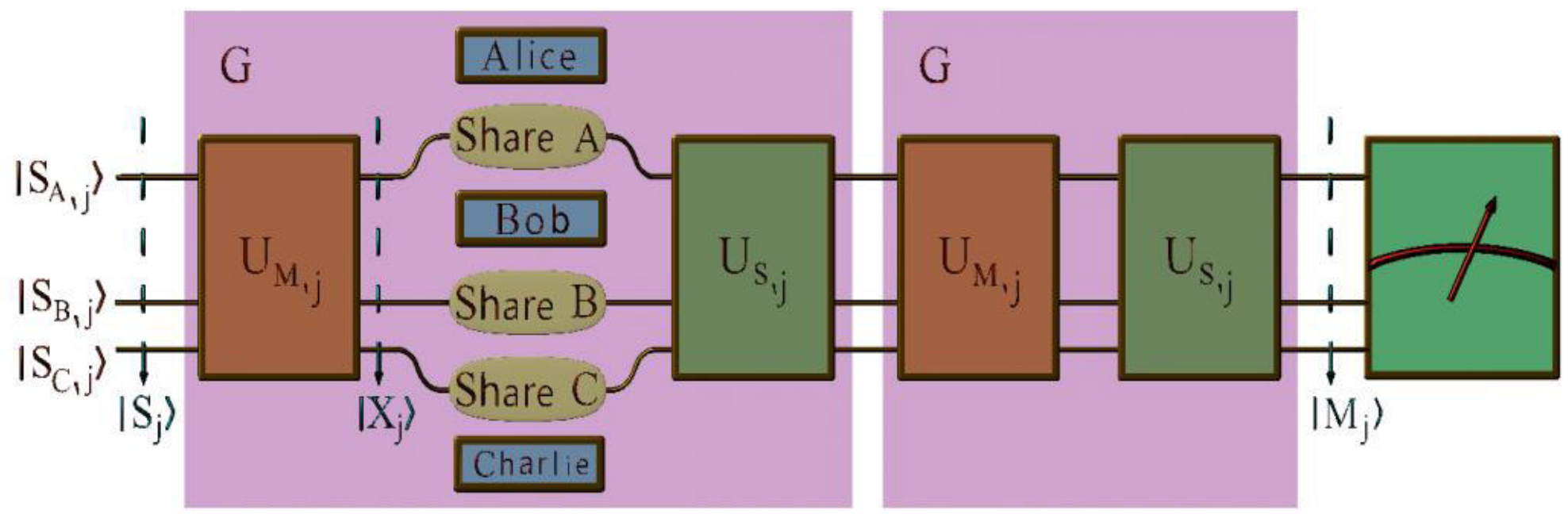

*Fig.4. Scheme of the secret sharing protocol in case of 3 participants. Encryption ends with applying of $U_{M,j}$. Decryption consist of operators $U_{S,j}U_{M,j}U_{S,j}$ that need to be done by a padlock.*

The probability of correctly decoding the shared secret (assuming there are no protocol violations) is approximately 0.945313.

### 4.3. Interception attack

If an eavesdropper intercepts the message, the distributor is notified by at least one of the participants and terminates the protocol. The worst possible case is when Eve can use the "padlock". In this case to obtain the secret, she only needs to make an assumption for the initial state:

$$\begin{aligned} |S'_j\rangle_4 &= |S'_A\rangle_2 \otimes |S'_B\rangle_2 && \textit{in the case of two participants} \\ |S'_j\rangle_8 &= |S'_A\rangle_2 \otimes |S'_B\rangle_2 \otimes |S'_C\rangle_2 && \textit{in the case of three participants} \end{aligned} \tag{31}$$

In order to construct the remaining part of the Grover's iteration Eve need to use the following reflection operator:

$$\begin{aligned} U'_{S,j} &= I_4 - 2|S'_j\rangle_4\langle S'_j|_4 && \textit{in the case of two participants} \\ U'_{S,j} &= I_8 - 2|S'_j\rangle_8\langle S'_j|_8 && \textit{in the case of three participants} \end{aligned} \tag{32}$$

Eve use $U'_S$ operator to complete the iteration:

$$\begin{aligned} |Z_j\rangle_4 &= U'_{S,j}|X_j\rangle_4 && \textit{in the case of two participants} \\ |Z_j\rangle_8 &= U'_{S,j}U_{M,j}U'_{S,j}|X_j\rangle_8 && \textit{in the case of three participants} \end{aligned} \tag{33}$$

Two examples are shown below:

1) In the case of two participants, secret message $M_j = 2$, initial state $|S_j\rangle_4 = |{+i}\rangle_2|+\rangle_2$ and assumed initial state $|S'_j\rangle_4 = |+\rangle_2|{-i}\rangle_2$, we have the secret are:

$$|Z_j\rangle_4 = \frac{1}{2}(|0\rangle_4 + |1\rangle_4 - i|2\rangle_4 + i|3\rangle_4) \tag{34}$$

The probability of obtaining the secret after measurement is:

$$P_M = 0.25 \tag{35}$$

2) For three participants, secret message $M_j = 1$, initial state $\left|S_j\right\rangle_8 = |+i\rangle_2 |+\rangle_2 |-\rangle_2$ and assumed initial state $\left|S'_j\right\rangle_8 = |+i\rangle_2 |+\rangle_2 |+i\rangle_2$ we have:

$$\left|X_j\right\rangle_8 = \frac{1}{2\sqrt{2}}\left(-|0\rangle_8 - |1\rangle_8 - |2\rangle_8 + |3\rangle_8 - i|4\rangle_8 + i|5\rangle_8 - i|6\rangle_8 + i|7\rangle_8\right) \tag{36}$$

$$\begin{aligned}\left|Z_j\right\rangle_8 = \frac{1}{2\sqrt{2}}\Bigg(&-\left(\frac{1}{2} - \frac{3i}{4}\right)|0\rangle_8 + \left(\frac{5}{4} - \frac{3i}{2}\right)|1\rangle_8 - \left(\frac{1}{2} - \frac{3i}{4}\right)|2\rangle_8 + \left(\frac{1}{4} + \frac{i}{2}\right)|3\rangle_8 \\ &- \left(\frac{3}{4} + \frac{i}{2}\right)|4\rangle_8 - \left(\frac{1}{2} - \frac{i}{4}\right)|5\rangle_8 - \left(\frac{3}{4} + \frac{i}{2}\right)|6\rangle_8 - \left(\frac{1}{2} - \frac{i}{4}\right)|7\rangle_8\Bigg)\end{aligned} \tag{37}$$

The probability of obtaining the secret after measuring $\left|Z_j\right\rangle_8$ is:

$$\mathrm{P_M} = 0.476563 \tag{38}$$

It is also possible that the eavesdropper is one of the participants. As in the case of external eavesdropping, at least one of the participants reports not receiving their respective qubit, and the distributor does not send the initial states. Thus there is no change in the considerations presented here.

## 5. Probability to crack the protocol by an interception attack - analytical solutions:

In this section, we provide an exact analytical solution of the probability for the eavesdropper to obtain the secret via an interception attack, in the case of two and three participants. In this study we replace Householder reflections with generalized Householder reflections with same phases to construct Grover's iterations.

### 5.1. Analytical solution for two participants and an arbitrary phase

Suppose the secret message $M$ is an integer in the interval $[0,3]$. It can be encoded as a two qubit state number between 0 and 3. The analysis bellow can be repeated for each two qubit group.

The secret message $|M\rangle_4$ can be written as:

$$|M\rangle_4 = \sum_{j=0}^{3} \delta_{j,M} |j\rangle_4 \tag{39}$$

The protocol begins with Dan (D) preparing two qubits (one for each of the two participants: Alice (A) and Bob (B)) in the following initial states:

$$\begin{aligned}|S_A\rangle_2 &= \left(|0\rangle_A + e^{i\varphi_1}|1\rangle_A\right)/\sqrt{2} \\ |S_B\rangle_2 &= \left(|0\rangle_B + e^{i\varphi_2}|1\rangle_B\right)/\sqrt{2}\end{aligned} \tag{40}$$

where $\varphi_1, \varphi_2 \in \left\{0, \frac{\pi}{2}, \pi, \frac{3\pi}{2}\right\}$. Each of the initial states can be expressed using these phases:

$$|S_A\rangle_2 = \begin{cases} |+\rangle_A = \left(|0\rangle_A + |1\rangle_A\right)/\sqrt{2} & \varphi_1 = 0 \\ |-\rangle_A = \left(|0\rangle_A - |1\rangle_A\right)/\sqrt{2} & \varphi_1 = \pi \\ |+i\rangle_A = \left(|0\rangle_A + i|1\rangle_A\right)/\sqrt{2} & \varphi_1 = \pi/2 \\ |-i\rangle_A = \left(|0\rangle_A - i|1\rangle_A\right)/\sqrt{2} & \varphi_1 = 3\pi/2 \end{cases} \tag{41}$$

Same goes for Bob. Thus, the initial state of the protocol is:

$$|S\rangle_4 = |S_A\rangle_2 \otimes |S_B\rangle_2 = \frac{1}{2}\sum_{j=0}^{3} i^{\alpha_j}\,|j\rangle_4 \tag{42}$$

where

$$\alpha_j = \begin{cases} 0 & j=0 \\ 2\varphi_2/\pi & j=1 \\ 2\varphi_1/\pi & j=2 \\ 2(\varphi_1+\varphi_2)/\pi & j=3 \end{cases} \tag{43}$$

The distributor D encodes the message by applying a generalized Householder reflection operator $U_M$, constructed using the state $|M\rangle_4$ and a phase $\Omega$. The phase that should be used to obtain the secret with certainty is $\Omega_{MAX} = \pi$.

$$U_M = I_4 - \left(1 - e^{i\Omega}\right)|M\rangle_4\langle M|_4 \tag{44}$$

The state $\left|X_j\right\rangle_4$ distributed among the participants is:

$$|X\rangle_4 = U_{M,j}|S\rangle_4 = |S\rangle_4 - \frac{\left(1-e^{i\Omega}\right)}{2} i^{\alpha_M}|M\rangle_4 \tag{45}$$

where $\alpha_M$ is $\alpha_j$, with $j = M$.

Suppose an eavesdropper Eve (E) intercepts the message. At least one of the participants (e.g. A) tells D that have not received their respective qubit from the secret. The distributor (D) terminates the protocol and does not disclose the bases of the initial qubit states $|S_A\rangle_2$ and $|S_B\rangle_2$.

The eavesdropper has the following possible strategies:

1) To take an arbitrary number that can be encoded in two qubits. The probability of guessing correctly is $P_G = 1/4$.
2) To try to complete the protocol procedure. In this case she makes an assumption for the initial state of each qubit (4 possibilities for each qubit):

$$\begin{aligned} |S'_A\rangle_2 &= \left(|0\rangle_A + e^{i\varphi'_1}|1\rangle_A\right)/\sqrt{2} \\ |S'_B\rangle_2 &= \left(|0\rangle_B + e^{i\varphi'_2}|1\rangle_B\right)/\sqrt{2} \end{aligned} \tag{46}$$

The bases $\{|+\rangle_A, |-\rangle_A\}$ and $\{|+i\rangle_A, |-i\rangle_A\}$ are mutually unbiased:

$$\langle S'_A|S_A\rangle = \frac{1+e^{i(\varphi_1-\varphi'_1)}}{2} \qquad \langle S'_B|S_B\rangle = \frac{1+e^{i(\varphi_2-\varphi'_2)}}{2} \tag{47}$$

We have:

$$|\langle S'_A|S_A\rangle|^2 = \begin{cases} 1 & \left|\varphi_1-\varphi'_1\right| = 0 \\ 0 & \left|\varphi_1-\varphi'_1\right| = \pi \\ 1/2 & \left|\varphi_1-\varphi'_1\right| = \pi/2 \end{cases} \tag{48}$$

$$|\langle S'_B|S_B\rangle|^2 = \begin{cases} 1 & \left|\varphi_2-\varphi'_2\right| = 0 \\ 0 & \left|\varphi_2-\varphi'_2\right| = \pi \\ 1/2 & \left|\varphi_2-\varphi'_2\right| = \pi/2 \end{cases}$$

Eve's assumption for the initial state $\left|S'\right\rangle_4$ can be expressed as:

$$|S'\rangle_4 = |S'_A\rangle_2 \otimes |S'_B\rangle_2 = \frac{1}{2}\sum_{j=0}^{3} i^{\alpha'_j}\,|j\rangle_4 \tag{49}$$

Where:

$$\alpha'_j = \begin{cases} 0 & j = 0 \\ 2\varphi'_2/\pi & j = 1 \\ 2\varphi'_1/\pi & j = 2 \\ 2(\varphi'_1 + \varphi'_2)/\pi & j = 3 \end{cases} \tag{50}$$

She does not know Dan's initial state $|S\rangle_4$. So instead of $|S\rangle_4$ she constructs operator $U'_S$ using her chosen $|S'\rangle_4$.

$$U'_S = I - \left(1 - e^{i\Omega}\right)|S'\rangle_4\langle S'|_4 \tag{51}$$

Grover's algorithm in this case of a two-cube register consists of only one iteration, so she applies $U'_S$ to complete protocol procedure:

$$\begin{aligned} |Z\rangle_4 = U'_S|X\rangle_4 = \\ |S\rangle_4 - \left(1 - e^{i\Omega}\right)|S'\rangle_4\langle S'|S\rangle_4 - \frac{\left(1 - e^{i\Omega}\right)}{2} i^{\alpha_M}|M\rangle_4 \\ + \frac{\left(1 - e^{i\Omega}\right)^2}{2} i^{\alpha_M}|S'\rangle_4\langle S'|M\rangle_4 \end{aligned} \tag{52}$$

After that Eve measures the state of $|Z\rangle_4$. The probability for her to obtain the secret is:

$$P_M(\Omega, S, S', M) = |\langle M|Z\rangle|^2 = \left|-\frac{\left(1 - e^{i\Omega}\right)}{2} i^{\alpha'_M}\langle S'|S\rangle_4 + \frac{i^{\alpha_M}}{8}\left(1 + e^{i\Omega}\right)^2\right|^2 \tag{53}$$

The analytical formula for the probability of obtaining the secret can also be expressed as follows:

$$\begin{aligned} P_M\left(\Omega, \varphi_1, \varphi'_1, \varphi_2, \varphi'_2, 0\right) \\ = \frac{1}{64}\Big|3e^{i\Omega} + e^{-2i\Omega} + e^{i(\varphi_1 - \varphi'_1)}\left(e^{i\Omega} - 1\right) + e^{i(\varphi_2 - \varphi'_2)}\left(e^{i\Omega} - 1\right) \\ + e^{i(\varphi_1 + \varphi_2 - (\varphi'_1 + \varphi'_2))}\left(e^{i\Omega} - 1\right)\Big|^2 \end{aligned} \tag{54}$$

$$\begin{aligned} P_M\left(\Omega, \varphi_1, \varphi'_1, \varphi_2, \varphi'_2, 1\right) \\ = \frac{1}{64}\Big|e^{i(\varphi_1 + \varphi_2 - \varphi'_1)}\left(e^{i\Omega} - 1\right) + e^{i(\varphi_1 + \varphi'_1 - \varphi'_2)}\left(e^{i\Omega} - 1\right) \\ + e^{i\varphi'_2}\left(e^{i\Omega} - 1\right) + 3e^{i\varphi_2 - 2i\Omega}\left(1/3 + e^{i\Omega}\right)\Big|^2 \end{aligned} \tag{55}$$

$$\begin{aligned} P_M\left(\Omega, \varphi_1, \varphi'_1, \varphi_2, \varphi'_2, 2\right) \\ = \frac{1}{64}\Big|e^{i(\varphi_1 + \varphi_2 - \varphi'_2)}\left(e^{i\Omega} - 1\right) + e^{i(\varphi_2 + \varphi'_1 - \varphi'_2)}\left(e^{i\Omega} - 1\right) \\ + e^{i\varphi'_1}\left(e^{i\Omega} - 1\right) + 3e^{i\varphi_1 - 2i\Omega}\left(1/3 + e^{i\Omega}\right)\Big|^2 \end{aligned} \tag{56}$$

$$\begin{aligned} P_M\left(\Omega, \varphi_1, \varphi'_1, \varphi_2, \varphi'_2, 3\right) \\ = \frac{1}{64}\Big|-e^{i\left(\varphi_2 - \Omega + \varphi'_1\right)}\left(e^{i\Omega} - 1\right) - e^{i\left(\varphi_1 - \Omega + \varphi'_2\right)}\left(e^{i\Omega} - 1\right) \\ + e^{i\left(\varphi'_1 + \varphi'_2\right)}\left(e^{i\Omega} - 1\right) + 3e^{i(\varphi_1 + \varphi_2 - 2\Omega)}\left(1/3 + e^{i\Omega}\right)\Big|^2 \end{aligned} \tag{57}$$

Using Equation (53) (or alternatively Equations (54), (55), (56) and (57)), analytical results can be obtained for the probability $P_M$ of obtaining the secret as a function of the number of mistaken phases $\varphi_j$ (j=1,2) and the difference between the assumed and the actual values of these phases. For $\Omega = \pi$ Equation (53) simplifies to:

$$P_M(\pi, S, S', M) = |\langle M|Z\rangle_4|^2 = \left|\langle S'|S\rangle_4\right|^2 = \left|\langle S'_A|S_A\rangle_2\right|^2 \left|\langle S'_B|S_B\rangle_2\right|^2 \qquad (58)$$

The results of the evaluation are shown in *Table.2*. This table features the probability $P_M$ of obtaining the secret under various conditions, and the number $N_M$ of assumed states for which these conditions are fulfilled. The results do not depend on the assumed initial state, but only on the differences between the actual and the assumed initial state.

| № | Condition | $P_M(\Omega = \pi)$ | $N_M(\Omega = \pi)$ |
|---|---|---|---|
| 0 | $\left|\varphi'_j - \varphi_j\right| = 0\ \forall j \in \{1,2\}$ | 1 | 16 |
| 1 | $\left|\varphi'_j - \varphi_j\right| = \pi/2\ \&\ \left|\varphi'_k - \varphi_k\right| = 0\ \ k \neq j$ | 0.5 | 64 |
| 2 | $\left|\varphi'_1 - \varphi_1\right| = \pi/2\ \&\ \left|\varphi'_2 - \varphi_2\right| = \pi/2$ | 0.25 | 64 |
| 3 | $\exists\ j$ *such that* $\left|\varphi'_j - \varphi_j\right| = \pi$ | 0 | 112 |

*Table.2.Probability of obtaining the secret (third column) of Grover-based secret sharing protocol in the case of two participants as a function of the number of mistaken phases and the differences between the assumed and the actual phases (second column). The fourth column shows the number of combinations that fulfil these conditions respectively.*

*Fig.5* shows numerical simulations for the probability $P_M$ of obtaining the secret as a function of $|S\rangle_4$ and $|S'\rangle_4$. The numbers on the horizontal and vertical axes are the ones asigned to the vectors $|S\rangle_4$ and $|S'\rangle_4$ respectively. These numbers take values from *1* to *4ⁿ*, where n is the number of the participants, and are assigned according to the following rules:

1) The initial qubit states are assigned the following numbers: 0 for $|+\rangle_2$, 1 for $|-\rangle_2$, 2 for $|+i\rangle_2$ and 3 for $|-i\rangle_2$.
2) The numbers assigned to the states $|S\rangle_4$ and $|S'\rangle_4$ are evaluated according to the respective tensor product.

For example in the case of two participants the state $|+i\rangle_2|-\rangle_2$is assigned the number $21_{(4)}$ + 1 = 9 + 1 =10.

The plot a) corresponds to the optimal $\Omega = \Omega_{MAX} = \pi$ and b) to $\Omega = 1.33$. The legend is shown on their right side. These numerical results match the analytical shown in *Table.2*.

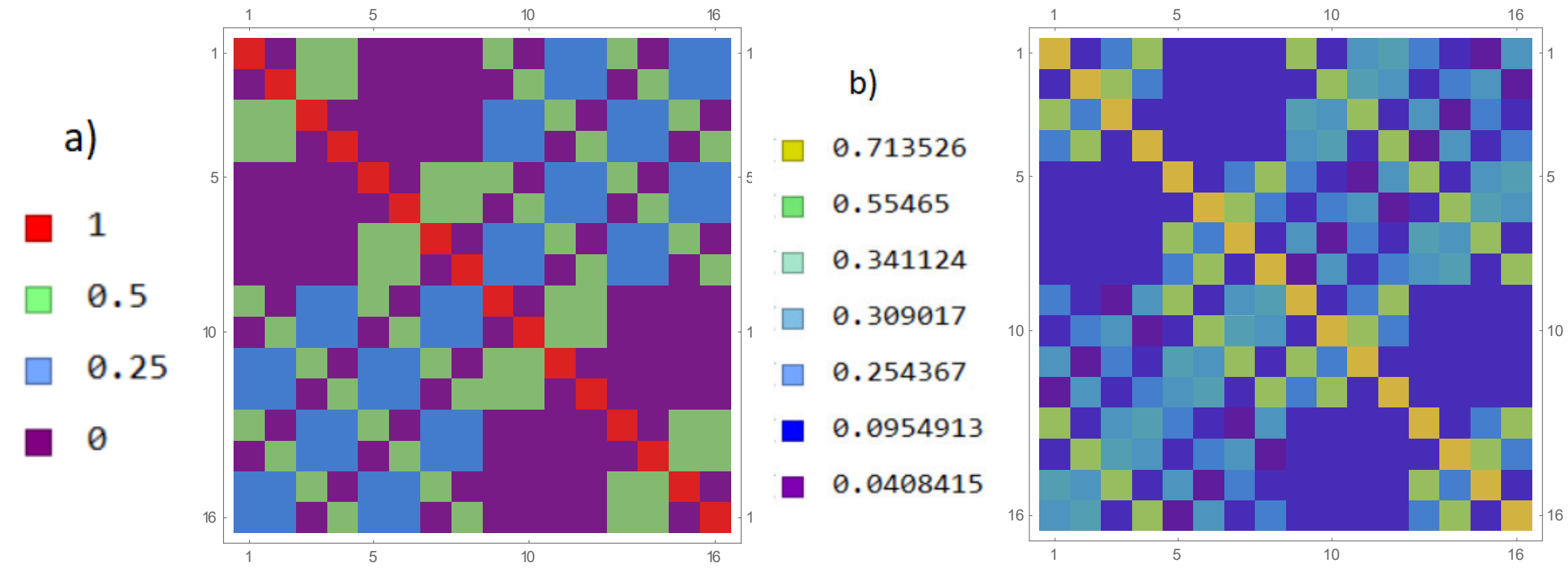

*Fig.5. Probability of obtaining solution of the secret sharing protocol based on Grover's algorithm in case of two participants for the angles* $\Omega = \pi$ *(plot a)) and* $\Omega = 1.33$ *(plot b)). Each row corresponds to the initial state S, and each column to the assumed initial state* $S'$. The *different probabilities of finding a solution are represented by different colours.*

Summing up the probabilities on each column gives the overall probability of obtaining the secret for the corresponding initial state $|S\rangle_4$. The probability evaluated for each row is the same, so the overall probability of obtaining the secret can be evaluated by using different probability values $P_M$ and the number $N_M$ of possible assumed states of $P_M$. Examples for $\Omega = \pi$ and $\Omega = 1.33$ are shown bellow:

$$P_S(\Omega = \pi) = \sum_M \frac{P_M(\pi) N_M(\pi)}{16^2} = 0.25 = P_G \tag{59}$$

$$P_S(\Omega = 1.33) = \sum_M \frac{P_M(1.33) N_M(1.33)}{16^2} = 0.25 = P_G \tag{60}$$

The probability for the eavesdropper to obtain the secret is 0.25 and does not depend on $\Omega$. However, $\Omega$ changes the probability that honest participants obtain the secret at the end of the protocol (see *Fig.5*).

The results of this chapter show that the probability for an eavesdropper to decode the secret message by intercepting the qubits sent to the other participants is the same as by random guessing.

### 5.2. Analytical solutions for three participants and an arbitrary phase

Suppose the secret message M is an integer in the interval $\{0,7\}$. It can be encoded as a three qubit state (if the secret message is a larger integer, the calculations will be applied to each set of three qubits). Without loss of generality, let the secret D wants to share be:

$$|M\rangle_8 = \sum_{j=0}^{7} \delta_{j,M} |j\rangle_8 \tag{61}$$

Dan (D) prepares three qubits (one for each of the participants: Alice (A), Bob (B) and Charlie (C)) in the following initial states:

$$\begin{aligned} |S_A\rangle_2 &= \left(|0\rangle_A + e^{i\varphi_1}|1\rangle_A\right)/\sqrt{2} \\ |S_B\rangle_2 &= \left(|0\rangle_B + e^{i\varphi_2}|1\rangle_B\right)/\sqrt{2} \\ |S_C\rangle_2 &= \left(|0\rangle_C + e^{i\varphi_3}|1\rangle_C\right)/\sqrt{2} \end{aligned} \tag{62}$$

It should be noted that with suitable phase all angles in the original algorithm can be obtained. An example of Alice's qubit can be seen in Eq. (41). The same applies to all other participants. Then the initial state of the protocol $|S\rangle_8$ is:

$$
\begin{aligned}
|S\rangle_8 = |S_A\rangle_2 \otimes |S_B\rangle_2 \otimes |S_C\rangle_2 \\
&= \frac{1}{2\sqrt{2}}\left(|0\rangle_8 + e^{i\varphi_3}|1\rangle_8 + e^{i\varphi_2}|2\rangle_8 + e^{i(\varphi_3+\varphi_2)}|3\rangle_8 + e^{i\varphi_1}|4\rangle_8\right. \\
&\left.+ e^{i(\varphi_3+\varphi_1)}|5\rangle_8 + e^{i(\varphi_2+\varphi_1)}|6\rangle_8 + e^{i(\varphi_3+\varphi_2+\varphi_1)}|7\rangle_8\right) \\
&= \frac{1}{2\sqrt{2}}\sum_{j=0}^{7} i^{\alpha_j}\,|j\rangle_8
\end{aligned}
\tag{63}
$$

Where:

$$
\alpha_j = \begin{cases}
0 & j = 0 \\
2\varphi_3/\pi & j = 1 \\
2\varphi_2/\pi & j = 2 \\
2(\varphi_3+\varphi_2)/\pi & j = 3 \\
2\varphi_1/\pi & j = 4 \\
2(\varphi_3+\varphi_1)/\pi & j = 5 \\
2(\varphi_2+\varphi_1)/\pi & j = 6 \\
2(\varphi_3+\varphi_2+\varphi_1)/\pi & j = 7
\end{cases}
\tag{64}
$$

There are two ways to construct a secret sharing protocol based on Grover's algorithm, based on the iteration during which the qubits are distributed among the participants.

#### 5.2.1. Distributing the qubits during the first iteration

The variant is described in [28] (only for the case when standard Householder reflections are used). The distributor encodes the message $M$ by applying a generalized Householder reflection operator $U_M$ on the initial state:

$$
U_M = I - \left(1 - e^{i\Omega}\right)|M\rangle_8\langle M|_8 \tag{65}
$$

The encoded state is:

$$
|X\rangle_8 = U_M|S\rangle_8 = |S\rangle_8 - \frac{\left(1 - e^{i\Omega}\right)}{2\sqrt{2}}\, i^{\alpha_M}|M\rangle_8 \tag{66}
$$

where $\Omega$ is the generalized Householder reflection's phase.

If the eavesdropper Eve (E) intercepts the message, the honest participant tells D that he has not received a qubit from the secret. The distributor (D) terminates the protocol and does not disclose the initial states of the participants' qubits.

Similarly to the case with two participants, the eavesdropper has different options:

1) To randomly choose one of the possible states of the algorithm and check if it is a solution. The probability of guessing correctly is $P_G = 0.125$.
2) To try to guess the initial state and complete the protocol procedure (if she has access to the padlock).
3) To put a specially prepared initial state into the padlock, if she has access to it.

4) To complete only the current iteration and measure the state.
5) To guess both the initial state and take a random oracle state (if she does not have access to the padlock).

In the following subsections, we provide analytical solutions only for option 2. Some evaluations and simulations for the other options are shown in the Appendix.

### 5.2.2. Security against interception when distribution is during the first iteration

Suppose Eve has access to the "padlock" and decides to complete the protocol procedure. Let her assumption for the initial state $|S'\rangle_8$ be

$$|S'\rangle_8 = |S'_A\rangle_2 \otimes |S'_B\rangle_2 \otimes |S'_C\rangle_2 = \frac{1}{2\sqrt{2}} \sum_{j=0}^{7} i^{a'_j} \, |j\rangle_8 \tag{67}$$

where:

$$\begin{aligned} |S'_A\rangle_2 &= \left(|0\rangle_A + e^{i\varphi'_1}|1\rangle_A\right)/\sqrt{2} \\ |S'_B\rangle_2 &= \left(|0\rangle_B + e^{i\varphi'_2}|1\rangle_B\right)/\sqrt{2} \\ |S'_C\rangle_2 &= \left(|0\rangle_C + e^{i\varphi'_3}|1\rangle_C\right)/\sqrt{2} \end{aligned} \tag{68}$$

Eve tries to decode the message following the protocol procedure by putting the assumed initial state into the padlock. The padlock is constructed using the operator

$$U'_S = I - (1 - e^{i\Omega})|S'\rangle_8\langle S'|_8 \tag{69}$$

instead of the operator that Alice, Bob and Charlie have constructed by using the actual initial state $|S\rangle_8$:

$$U_S = I - (1 - e^{i\Omega})|S\rangle_8\langle S|_8 \tag{70}$$

The register state after each operator is applied are as follows:

$$U'_S|X\rangle_8 = |S\rangle + \frac{(1 - e^{i\Omega})^2 i^{a_M - a'_M}}{8}|S'\rangle - \frac{(1 - e^{i\Omega}) i^{a_M}}{2\sqrt{2}}|M\rangle - (1 - e^{i\Omega})|S'\rangle\langle S'|S\rangle \tag{71}$$

$$\begin{aligned} U_M U'_S|X\rangle_8 = {} & |S\rangle + \frac{(1 - e^{i\Omega})^2 i^{a_M - a'_M}}{8}|S'\rangle + \frac{(1 - e^{i\Omega})^2 i^{a'_M}}{2\sqrt{2}}|M\rangle\langle S'|S\rangle \\ & - \frac{(1 - e^{i\Omega})(3 + e^{i\Omega})^2 i^{a_M}}{16\sqrt{2}}|M\rangle - (1 - e^{i\Omega})|S'\rangle\langle S'|S\rangle \end{aligned} \tag{72}$$

The padlock then completes the protocol procedure:

$$\begin{aligned} |Z\rangle_8 = {} & U'_S U_M U'_S|X\rangle_8 \\ & = -\frac{(1 - e^{i\Omega})(3 + e^{i\Omega})^2}{8}|S'\rangle\langle S'|S\rangle \\ & + \frac{(1 - e^{i\Omega})^2(9 + 14e^{i\Omega} + e^{2i\Omega})}{64} i^{a_M - a'_M}|S'\rangle + |S\rangle \\ & + \frac{(1 - e^{i\Omega})^2 i^{a'_M}}{2\sqrt{2}}|M\rangle\langle S'|S\rangle - \frac{(1 - e^{i\Omega})(3 + e^{i\Omega})^2 i^{a_M}}{16\sqrt{2}}|M\rangle \end{aligned} \tag{73}$$

Eve measures the state of $|Z\rangle_8$ after the end of the protocol.

$$P_M(\Omega, S, S', M) = |\langle M|Z\rangle_8|^2 = \left| -\frac{(1-e^{i\Omega})(1+14e^{i\Omega}+e^{2i\Omega})}{16\sqrt{2}} i^{a'_M - a_M} \langle S'|S\rangle + \frac{1+20e^{i\Omega}+22e^{2i\Omega}+20e^{3i\Omega}+e^{4i\Omega}}{128\sqrt{2}} \right|^2 \tag{74}$$

The probability of decoding the message is:

$$P_M(\Omega, \varphi_1, \varphi'_1, \varphi_2, \varphi'_2, \varphi_3, \varphi'_3, M) = 1.52587890.10^{-5} \left| V[\Omega] e^{iU[M]} + W[\Omega] \prod_{j=1}^{3} \left( e^{i\varphi_j} + e^{i\varphi'_j} \right) \right|^2 \tag{75}$$

where:

$$V[\Omega] = e^{6i\Omega}(1.41421 + 31.1126\, e^{i\Omega} + 67.8822\, e^{2i\Omega} - 8.48528\, e^{3i\Omega} - 1.41421 e^{4i\Omega}) \tag{76}$$

$$W[\Omega] = e^{7i\Omega}\left(1.41421 + 18.3847 e^{i\Omega} - 18.3847 e^{2i\Omega} - 1.41421 e^{3i\Omega}\right) \tag{77}$$

The function $U[M]$ is defined as shown in *Table 3*. The first and second columns give the integer in decimal and binary number systems, respectively. The third column gives the corresponding secret message, and the last column gives the value of the function $U[M]$. It should be noted that using the binary representation of the number, it is easy to guess the function.

| Message | Binary Representation | $\vert M\rangle$ | $U[M]$ |
|---|---|---|---|
| 0 | $000_2$ | $(0,0,0,0,0,0,0,1)^T$ | $\varphi'_1 + \varphi'_2 + \varphi'_3$ |
| 1 | $001_2$ | $(0,0,0,0,0,0,1,0)^T$ | $\varphi'_1 + \varphi'_2 + \varphi_3$ |
| 2 | $010_2$ | $(0,0,0,0,0,1,0,0)^T$ | $\varphi'_1 + \varphi_2 + \varphi'_2$ |
| 3 | $011_2$ | $(0,0,0,0,1,0,0,0)^T$ | $\varphi'_1 + \varphi_2 + \varphi_3$ |
| 4 | $100_2$ | $(0,0,0,1,0,0,0,0)^T$ | $\varphi_1 + \varphi'_2 + \varphi_3{}'$ |
| 5 | $101_2$ | $(0,0,1,0,0,0,0,0)^T$ | $\varphi_1 + \varphi'_2 + \varphi_3$ |
| 6 | $110_2$ | $(0,1,0,0,0,0,0,0)^T$ | $\varphi'_1 + \varphi'_2 + \varphi_3$ |
| 7 | $111_2$ | $(1,0,0,0,0,0,0,0)^T$ | $\varphi_1 + \varphi_2 + \varphi_3$ |

*Table 3. Values of the function $U[M]$ depending on the secret message $M$ encoded by the protocol.*

Analytical results for $\Omega = \Omega_{MAX} = 2.12688$ show that if Eve has guessed the initial state, the probability of obtaining the secret is equal to 1.

$$P_M(\pi, S, S', M) = |\langle M|Z\rangle_8|^2 = \left|\langle S'|S\rangle_8\right|^2 = \left|\langle S'_A|S_A\rangle_2\right|^2 \left|\langle S'_B|S_B\rangle_2\right|^2 \left|\langle S'_C|S_C\rangle_2\right|^2 \tag{78}$$

For each phase error of π/2, the probability of finding a solution decreases by a factor of two. If the difference in the assumption of one or more angles is equal to π, then the probability of finding a solution becomes zero. The table also shows the results for $\Omega = \pi$. *Table 4* shows how the probability $P_M$ of obtaining the secret depends on the differences between the actual and the assumed phase values in case of $\Omega = \pi$ and $\Omega = \Omega_{MAX}$. The table also gives the number of phase combinations $N_M$ that have probability $P_M$.

<table>
<tr><th>№</th><th>Condition</th><th>$P_M(\Omega=\pi)$</th><th>$N_M(\Omega=\pi)$</th><th>$P_M(\Omega=\Omega_{MAX})$</th><th>$N_M(\Omega=\Omega_{MAX})$</th></tr>
<tr><td>0</td><td>$\left|\varphi'_j-\varphi_j\right|=0$<br>$\forall j\in\{1,2,3\}$</td><td>0.9453</td><td>64</td><td>1</td><td>64</td></tr>
<tr><td>1</td><td>$\left|\varphi'_j-\varphi_j\right|=\pi/2$<br>$\left|\varphi'_k-\varphi_k\right|=0\forall\, k\neq j$</td><td>0.4766</td><td>384</td><td>0.5</td><td>384</td></tr>
<tr><td rowspan="2">2</td><td rowspan="2">$\left|\varphi'_j-\varphi_j\right|=\pi/2$<br>$\left|\varphi'_k-\varphi_k\right|=\pi/2 k\neq j$<br>$\left|\varphi'_l-\varphi_l\right|=0\forall l\neq j, l\neq k$</td><td>0.2891</td><td>384</td><td rowspan="2">0.25</td><td rowspan="2">768</td></tr>
<tr><td rowspan="2">0.1953</td><td rowspan="2">512</td></tr>
<tr><td rowspan="2">3</td><td rowspan="2">$\left|\varphi'_j-\varphi_j\right|=\pi/2$<br>$\forall j\in\{1,2,3\}$</td><td rowspan="2">0.125</td><td rowspan="2">512</td></tr>
<tr><td>0.1016</td><td>384</td></tr>
<tr><td>4</td><td>$\exists\, j$ such that $\left|\varphi'_j-\varphi_j\right|=\pi$</td><td>0.00781</td><td>2368</td><td>0</td><td>2368</td></tr>
</table>

*Table 4. The probability of obtaining the secret in the case of three participants depending on the differences between the assumed and the actual phase values.* $N_M$ *denotes the possible initial states fulfilling the respective conditions in case of* $\Omega=\Omega_{MAX}$.

The results of numerical simulations of $P_M$ as a function of the actual and the assumed initial state are shown in *Fig.6.* On the horizontal axis are the numbers assigned to the actual initial state $|S\rangle_8$ of the protocol and on the vertical axis are these assigned to the assumed state $|S'\rangle_8$.Where numbers are assigned to $|S\rangle_8$ and $|S'\rangle_8$ in a similar way to the case of two participants, for example the state $|+\rangle_2|-i\rangle_2|+i\rangle_2$ is assigned the number $032_{(4)}$ + 1 = 14 + 1 = 15. The plot a) corresponds to $\Omega=\Omega_{MAX}$, and the plot b) to $\Omega=\pi$. Each color corresponds to a different probability of finding a solution. The legend indicating the correspondence between the color and the probability of finding a solution is shown to the right of each picture

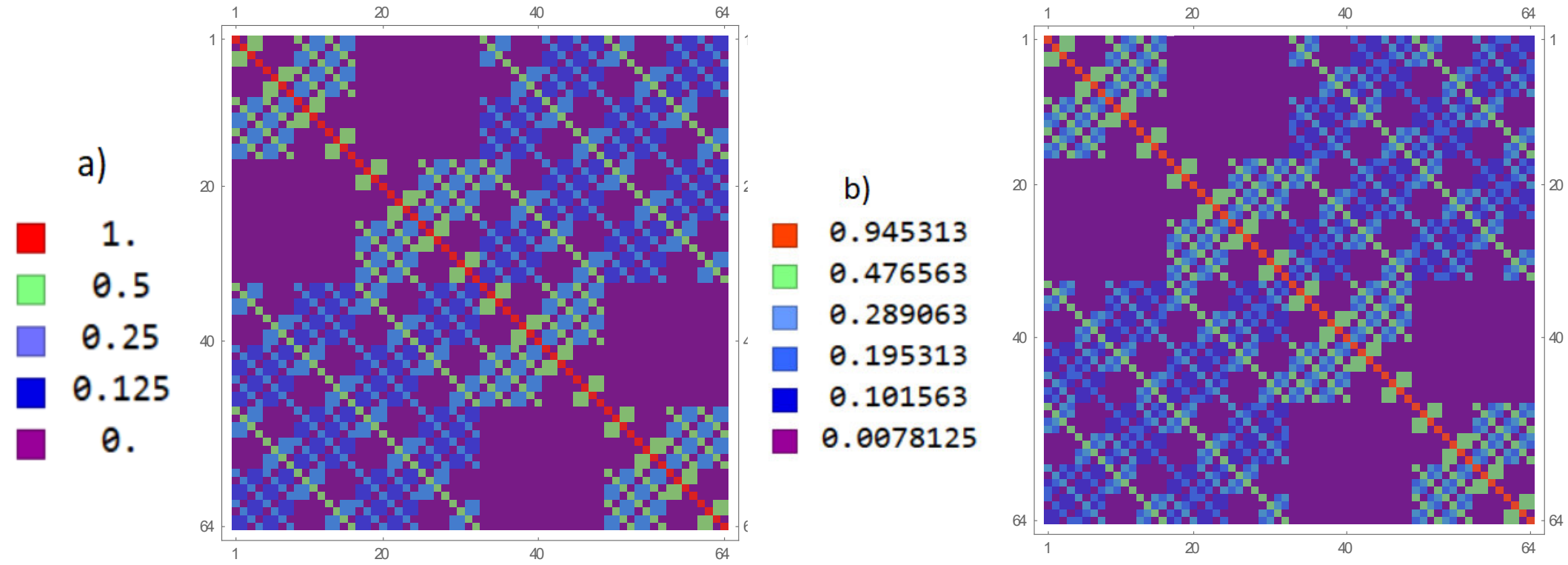

*Fig.6. Probability of an eavesdropper to obtain the secret via an interception attack in the case of three participants when the distribution of the qubits among the participants is done after applying the oracle in the first Grover iteration, as a function of the actual initial state* $|S\rangle_8$ *and the eavesdropper's assumption* $|S'\rangle_8$ *for the initial state, for* $\Omega=\Omega_{MAX}\approx 2.12688$ *(plot a)) and* $\Omega=\pi$ *(plot b)).*

The overall probability of finding the solution when trying to guess the initial state is:

$$P_S(\Omega=\Omega_{MAX})=\sum_M\frac{P_M(\Omega_{MAX})N_M(\Omega_{MAX})}{64^2}=\frac{1}{8}=P_G \qquad (79)$$

$$P_S(\Omega = \pi) = \sum\nolimits_M \frac{P_M(\pi)N_M(\pi)}{64^2} = \frac{1}{8} = P_G \tag{80}$$

### 5.2.3. Distribution during the second iteration. Interception attack analysis

In the second variant of the protocol the distributor Dan encodes the message $M$ by applying the first Grover iteration and then the oracle $U_S$ of the second iteration, thus obtaining the encoded state:

$$|Y\rangle_8 = U_M U_S U_M |S\rangle_8 \tag{81}$$

He then sends one qubit from the state to each of the participants - Alice, Bob and Charlie. In this case, $|Y\rangle_8$ can also be expressed as:

$$|Y\rangle_8 = {U_s}^{-1}|M\rangle_8 \tag{82}$$

where ${U_s}^{-1}$ is the inverse matrix of $U_s$.

In case there is no eavesdropper, each participant receives their respective qubit and reports receiving it. Dan then announces to them the initial states over the open channel and they apply $U_s$ to obtain the secret:

$$|M\rangle_8 = U_S|Y\rangle_8 \tag{83}$$

Note that the decryption is done without using the secret state $|M\rangle_8$, so unlike in the first variant, the secret does not need to be stored in a “padlock”.

The steps of the protocol can be depicted as shown in *Fig.7*.

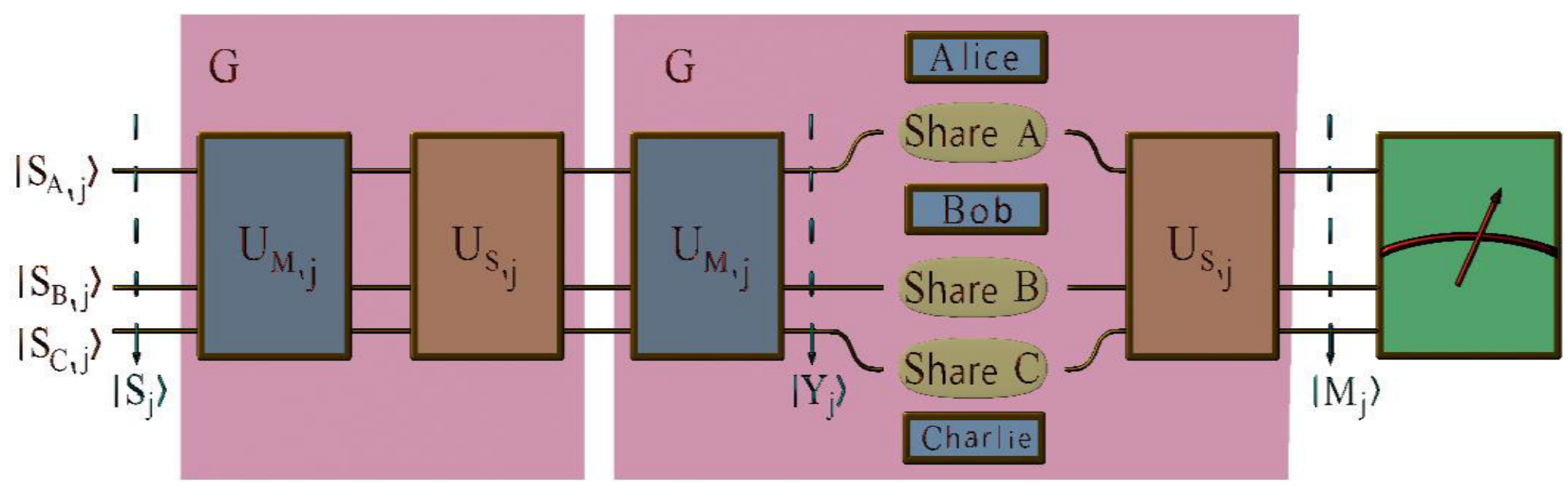

*Fig.7. Second variant of the three-party secret sharing protocol, in which distributing the qubits among the participants is done after applying the oracle in the second Grover iteration. The decryption can be done without using a "padlock".*

If an eavesdropper, Eve (E), intercepts the message $|Y\rangle_8$, the honest participant tells D that he has not received his qubit from the secret. The distributor (D) terminates the protocol and the initial states remains undisclosed.

The eavesdropper has only two possible options:

1) To randomly choose one of the possible states and check if it is a solution. The probability that she guesses it correctly is $P_G = 0.125$.
2) To try to guess the initial state and complete the protocol procedure. In this case, she constructs her assumption $|S'\rangle_8$ for the initial state (68) and the corresponding $U'_S$ gate (69).

She then applies $U'_S$ to the state $|Y\rangle_8$:

$$|Z\rangle_8 = U'_S|Y\rangle_8 \tag{84}$$

Then Eve measures the state of $|Z\rangle_8$.

Analytical results show that, the overall probability for the eavesdropper to obtain the secret is the same as in the first variant:

$$P_S(\Omega = \Omega_{MAX}) = \sum\nolimits_M \frac{P_M(\Omega_{MAX})N_M(\Omega_{MAX})}{64^2} = \frac{1}{8} = P_G \tag{85}$$

$$P_S(\Omega = \pi) = \sum\nolimits_M \frac{P_M(\pi)N_M(\pi)}{64^2} = \frac{1}{8} = P_G \tag{86}$$

This means that both variants of the protocol are secure against interception attacks.

## 6. Generalization of the protocol for any number of participants

In this section, we present various ways of generalizing the protocol for an arbitrary number of participants, and analyze its security against interception attacks.

### 6.1. Applying Grover's algorithm to the initial state $|S\rangle_N$

Let the initial state in the algorithm be:

$$|S\rangle_N = \frac{1}{\sqrt{N}}\sum_{j=0}^{N-1} e^{i\varphi_j}|j\rangle_N = e^{i\varphi_M}\left(\frac{1}{\sqrt{N}}|M\rangle_N + \sqrt{\frac{N-1}{N}}|M^\perp\rangle_N\right) = e^{i\varphi_M}|S_1\rangle_N \tag{87}$$

where:

$$|M^\perp\rangle_N = \frac{1}{\sqrt{N-1}}\sum_{\substack{j=0 \\ j\neq M}}^{N-1} e^{i(\varphi_j-\varphi_M)}|j\rangle_N \tag{88}$$

$$|S_1\rangle_N = \frac{1}{\sqrt{N}}|M\rangle_N + \sqrt{\frac{N-1}{N}}|M^\perp\rangle_N \tag{89}$$

The reflection operator $U_S$ can be expressed as:

$$U_S = \left(1-e^{i\omega}\right)|S\rangle_N\langle S|_N - I = \left(1-e^{i\omega}\right)|S_1\rangle_N\langle S_1|_N - I = U_{S_1} \tag{90}$$

Thus the Grover iteration $G$ can be written as

$$G = U_{S_1} U_M \tag{91}$$

Therefore, applying Grover's algorithm on $|S\rangle$ is equivalent to applying it to $|S_1\rangle$ and then multiplying the resulting state vector by a phase factor $e^{i\varphi_m}$. Thus, the state obtained at the end of the algorithm is

$$G^k|S\rangle_N = e^{i\varphi_M} G^k |S_1\rangle_N = e^{i\varphi_M} G^k \left( \frac{1}{\sqrt{N}} |M\rangle_N + \sqrt{\frac{N-1}{N}} |M^\perp\rangle_N \right) = e^{i\varphi_M} |M\rangle_N \tag{92}$$

This can be used to construct a quantum secret sharing protocol for arbitrary number of participants, similar to the ones described above for two and three participates. In the next two subsections we analyze the security of the protocol against interception attack in two cases: when the distribution of the qubits among the participants is done during the first or during the last Grover iteration respectively.

### 6.2. Distributing the qubits during the first Grover iteration

Consider the case where the distribution of the qubits among the participants is done after applying the oracle in the first Grover iteration (see *Fig.8*).

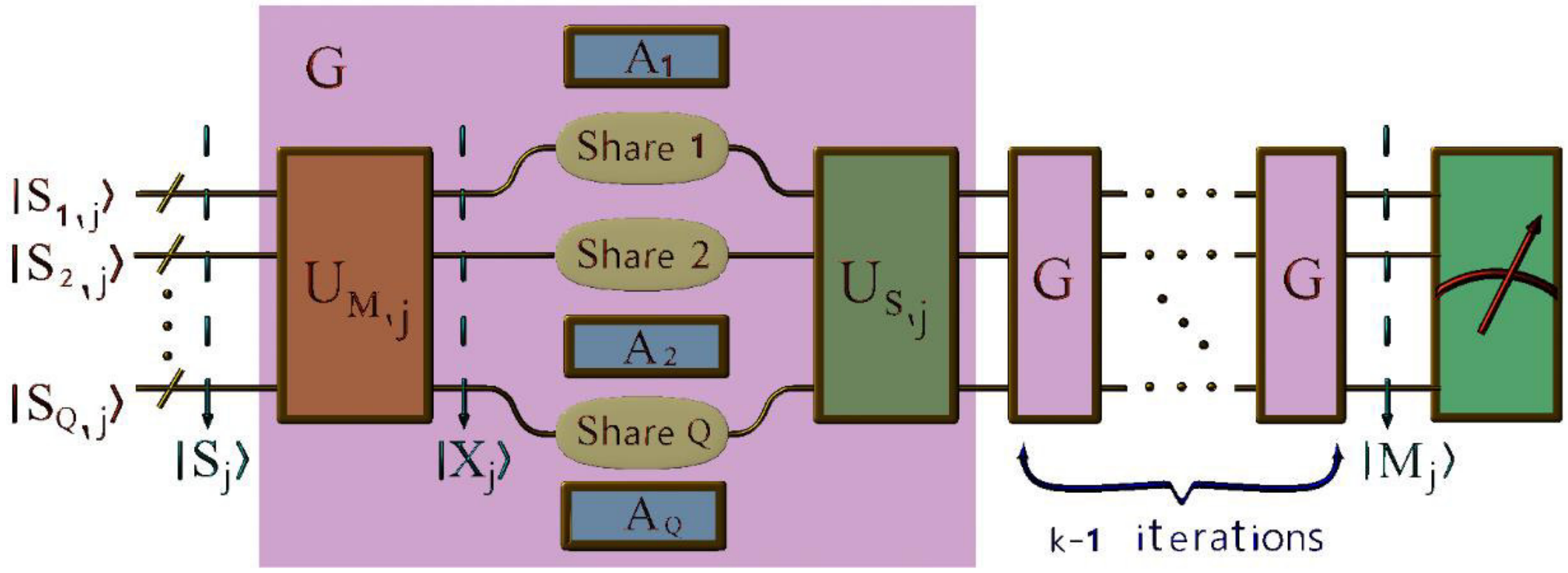

*Fig.8. Quantum secret sharing protocol for an arbitrary number of participants with qubits being distributed among the participants after the oracle in the first Grover iteration is applied.*

Suppose an eavesdropper, Eve, intercepts the participants' qubits. To finish the algorithm and obtain the secret key, she must first recover the $U_S$ gate, meaning that she needs to guess what the initial state $|S\rangle$ is. Let her assumption for the initial state be

$$|S'\rangle = \frac{1}{\sqrt{N}} \sum_{j=0}^{N-1} e^{i\varphi'_j} |j\rangle_N \tag{93}$$

She uses this state to construct the reflection operators:

$$U_{S'} = (1 - e^{i\omega})|S'\rangle\langle S'| - I \tag{94}$$

Her assumption for the Grover iteration is:

$$G' = U_{S'} U_M \tag{95}$$

Thus, finishing Grover's algorithm, Eve obtains the state

$$|M'\rangle = (G')^{k-1} U_{S'} (U_M |S\rangle) = (G')^{k-1} G' |S\rangle = (G')^k |S\rangle \tag{96}$$

State $|S'\rangle$ can also be expressed as

$$|S'\rangle = e^{i\varphi'_M} \left( \frac{1}{\sqrt{N}} |M\rangle_N + \sqrt{\frac{N-1}{N}} |M^{\perp\prime}\rangle \right) = e^{i\varphi'_M} |S'_1\rangle \tag{97}$$

where

$$|M^{\perp\prime}\rangle = \frac{1}{\sqrt{N-1}} \sum_{\substack{j=0 \\ j\neq M}}^{N-1} e^{i(\varphi'_j - \varphi'_M)} |j\rangle_N \tag{98}$$

$$|S'_1\rangle = \frac{1}{\sqrt{N}} |M\rangle_N + \sqrt{\frac{N-1}{N}} |M^{\perp\prime}\rangle \tag{99}$$

If the operator $G'$ is applied k times to $|S'\rangle$, the resulting state is

$$G'^k |S'\rangle = e^{i\varphi'_M} G'^k |S'_1\rangle = e^{i\varphi'_M} G'^k \left( \frac{1}{\sqrt{N}} |M\rangle_N + \sqrt{\frac{N-1}{N}} |M^{\perp\prime}\rangle \right) = e^{i\varphi'_M} |M\rangle_N \tag{100}$$

Hence

$$|M\rangle_N = e^{-i\varphi'_M} G'^k |S'\rangle \tag{101}$$

and the probability for Eve to obtain the secret key after measuring her state $|m'\rangle$ is

$$P = \left| \langle M|_N |M'\rangle \right|^2 = \left| e^{i\varphi'_M} \langle S'| (G'^k)^\dagger G'^k |S\rangle \right|^2 = |\langle S'|S\rangle|^2 \tag{102}$$

### 6.3. Distributing the qubits during the last Grover iteration

Suppose the distribution of the qubits among the participants is done after applying the oracle in the last Grover iteration (see *Fig.9*).

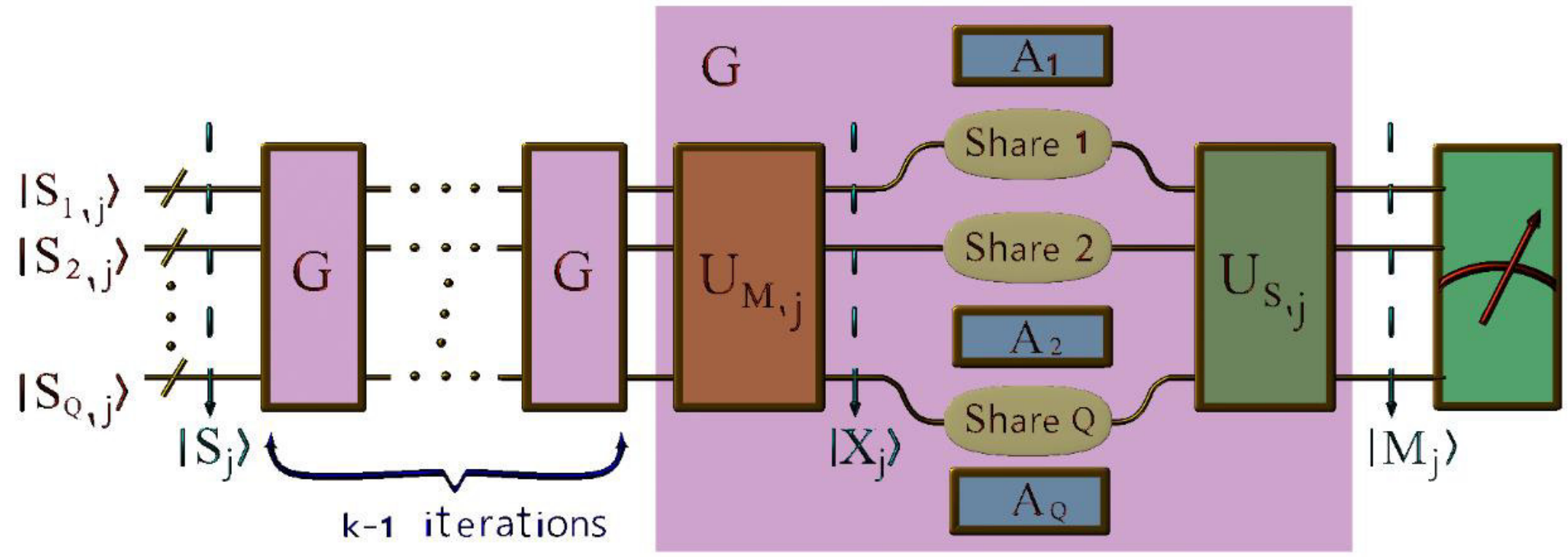

*Fig.9. Quantum secret sharing protocol for an arbitrary number of participants with qubits being distributed among the participants after the oracle in the last Grover iteration is applied. This variant does not require padlock.*

The result from Subsection 6.1, can be rewritten as:

$$\left|M_j\right\rangle_N = e^{-i\varphi_M}\left(G_j\right)^k\left|\mathrm{S}_j\right\rangle_N = e^{-i\varphi_M}U_{S,j}U_{M,j}\left(G_j(M,\varphi)\right)^{k-1}\left|\mathrm{S}_j\right\rangle_N \tag{103}$$

The state $\left|\mathrm{X}_j\right\rangle_N$ distributed among the participants is:

$$\left|\mathrm{X}_j\right\rangle_N = e^{-i\varphi_M}U_{M,j}G_j^{k-1}\left|\mathrm{S}_j\right\rangle_N = \left(U_{S,j}\right)^{-1}\left|M_j\right\rangle_N \tag{104}$$

If an eavesdropper, Eve, intercepts $|\mathrm{X}\rangle_N$, she makes an assumption for the initial state $|S'\rangle$:

$$|\mathrm{S}'\rangle_N = |S'_1\rangle_2 \otimes \ldots \otimes \left|S'_Q\right\rangle_2 \tag{105}$$

For the state $|M'\rangle_N$ obtained by Eve at the end of the protocol procedure:

$$\left|M_j{}'\right\rangle_N = U_{S',j}\left(U_{S,j}\right)^{-1}\left|M_j\right\rangle_N \tag{106}$$

The above equation can be written as:

$$\begin{aligned}\left|M_j{}'\right\rangle_N = \left|M_j\right\rangle_N - \left(1-e^{i\omega}\right)\left\langle \mathrm{S}_j{}'\middle|M_j\right\rangle_N\left|\mathrm{S}_j{}'\right\rangle_N - \left(1-e^{-i\omega}\right)\left\langle \mathrm{S}_j\middle|M_j\right\rangle_N\left|\mathrm{S}_j\right\rangle_N \\ + \left(2-e^{i\omega}-e^{-i\omega}\right)\left\langle S_j{}'\middle|S_j\right\rangle_N\left\langle \mathrm{S}_j\middle|M_j\right\rangle_N\left|\mathrm{S}_j{}'\right\rangle_N\end{aligned} \tag{107}$$

where:

$$U_{S,j} = I - \left(1-e^{i\Omega}\right)\left|S_j\right\rangle_{2^Q}\left\langle S_j\right|_{2^Q} \tag{108}$$

$$U_{S',j} = I - \left(1-e^{i\Omega}\right)\left|S'_j\right\rangle_{2^Q}\left\langle S'_j\right|_{2^Q} \tag{109}$$

To deduce the probability for Eve to obtain the secret, we will use following scalar product:

$$\left\langle \mathrm{M}_j\right|_N\left|M_j{}'\right\rangle_N = 1 - \frac{1}{N}\left(2-e^{i\omega}-e^{-i\omega}\right)\left(\left\langle S_j{}'\middle|S_j\right\rangle_N e^{i\left(\xi'_{M,j}-\xi_{M,j}\right)} - 1\right) \tag{110}$$

where we have made the following substitutions:

$$\langle M_j | S_j \rangle_N = e^{i\xi_{M,j}}/\sqrt{N} \qquad \langle M_j | S_j' \rangle_N = e^{i\xi'_{M,j}}/\sqrt{N} \tag{111}$$

Let the K-th qubit in the actual and assumed initial state be as follows:

$$|S_{K,j}\rangle_2 = \frac{|0\rangle + e^{i\varphi_{K,j}}|1\rangle}{\sqrt{2}} \qquad |S'_{K,j}\rangle_2 = \frac{|0\rangle + e^{i\varphi'_{K,j}}|1\rangle}{\sqrt{2}} \tag{112}$$

Their scalar products with the respective qubit $|m\rangle_N$ from the secret state are:

$$\langle M_{K,j}|_2 |S_{K,j}\rangle_N = \frac{e^{i\delta_{1,M_{K,j}}\varphi_{K,j}}}{\sqrt{2}} \qquad \langle M_{K,j}|_2 |S'_{K,j}\rangle_N = \frac{e^{i\delta_{1,M_{K,j}}\varphi'_{K,j}}}{\sqrt{2}} \tag{113}$$

This means that:

$$e^{i\xi_M} = e^{i\sum_{K=1}^{Q}\delta_{1,M_K}\varphi_K} \qquad e^{i\xi'_M} = e^{i\sum_{K=1}^{Q}\delta_{1,M_K}\varphi'_K} \tag{114}$$

For the scalar product $\langle S_K' | S_K \rangle_N$ we have:

$$\langle S_{K,j} | S'_{K,j} \rangle_2 = \frac{1}{2}\left(1 + e^{i(\varphi_{K,j} - \varphi'_{K,j})}\right) \tag{115}$$

This can be used to evaluate the product of true and guessed bases:

$$\langle S_j' |_N |S_j\rangle_N = \prod_{j=1}^{Q} \langle S_{K,j} | S'_{K,j} \rangle_2 = \left| \langle S_j' |_N |S_j\rangle_N \right| e^{i(\xi_M - \xi'_M)} \tag{116}$$

Thus the probability for Eve to obtain the secret after the final measurement is:

$$\left| \langle \mathrm{M} |_N | M' \rangle_N \right|^2 = \left| 1 - \frac{4}{N} sin^2\left(\frac{\omega}{2}\right) (|\langle S' | S \rangle_N| - 1) \right|^2$$

### 6.4. Distributing the qubits during another Grover iteration

In this variant of the protocol, the message is encoded by applying the Grover iteration to the initial state a certain number $k_1 - 1$ of times and then the oracle in the $k_1 - th$ iteration.

$$|X_j\rangle_{2^Q} = U_{M,j} G_j^{k_1 - 1} |S_j\rangle_{2^Q} \tag{117}$$

The integer $k_1$ is in the interval $[0, k-1]$, where k being the number of Grover iterations in the protocol:

One qubit of each part is then sent to each participant. When all of them have confirmed that they have received their respective qubit, the initial states of all qubits are sent over the open channel.

Decoding the message is done on the "padlock". This requires the qubits of all participants and the initial state of the system. The device then completes protocol procedure:

$$\left|M_j\right\rangle_{2Q} = \underbrace{G_j \ldots G_j}_{\mathrm{k}-\mathrm{k}_1\ times} U_{S,j}\left|X_j\right\rangle_{2Q} \tag{118}$$

This variant of the protocol is shown on *Fig.10*:

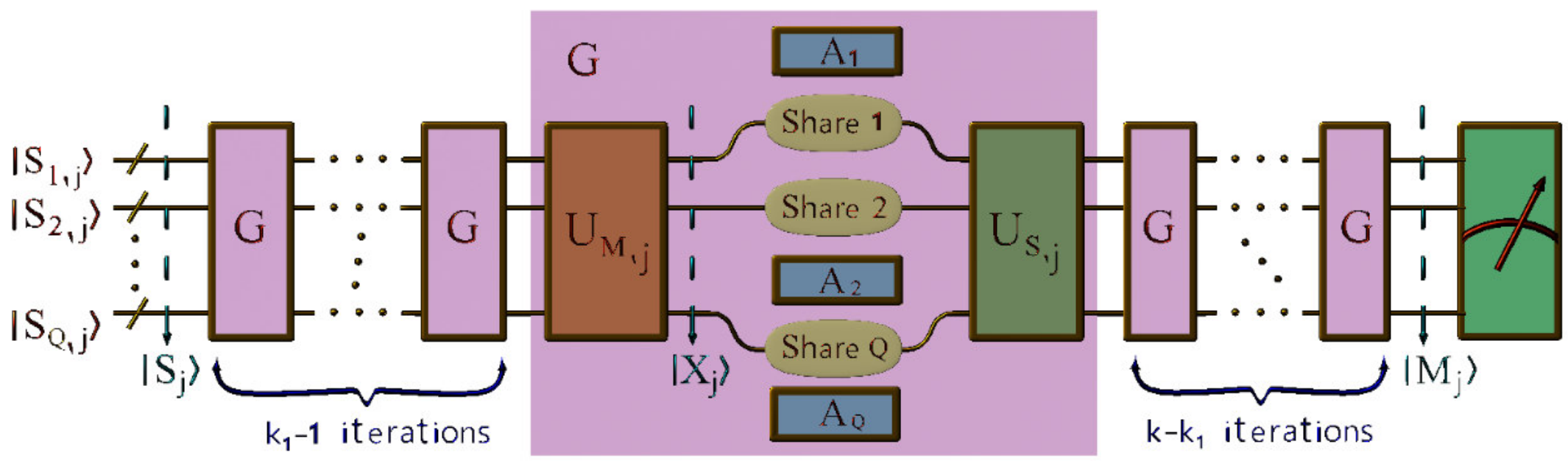

*Fig.10. Secret яharing protocol for an arbitrary number of participants with qubits being distributed between participants after the oracle from the* $\mathrm{k}_1 - th$ *iteration is applied.*

*Fig.11* shows the probability of obtaining the secret in the case of four participants. Numerical results for $k_1 = 0$, $k_1 = k - 1$ and $k_1 = \lfloor k/2 \rfloor - 1$, show shame dependence on the guessed states by Eve. The horizontal axis corresponds to the actual initial state and the vertical axis to the assumed initial state. The color corresponds to the probability to obtain the secret. Calculations are done for $\Omega = \Omega_{MAX}$.

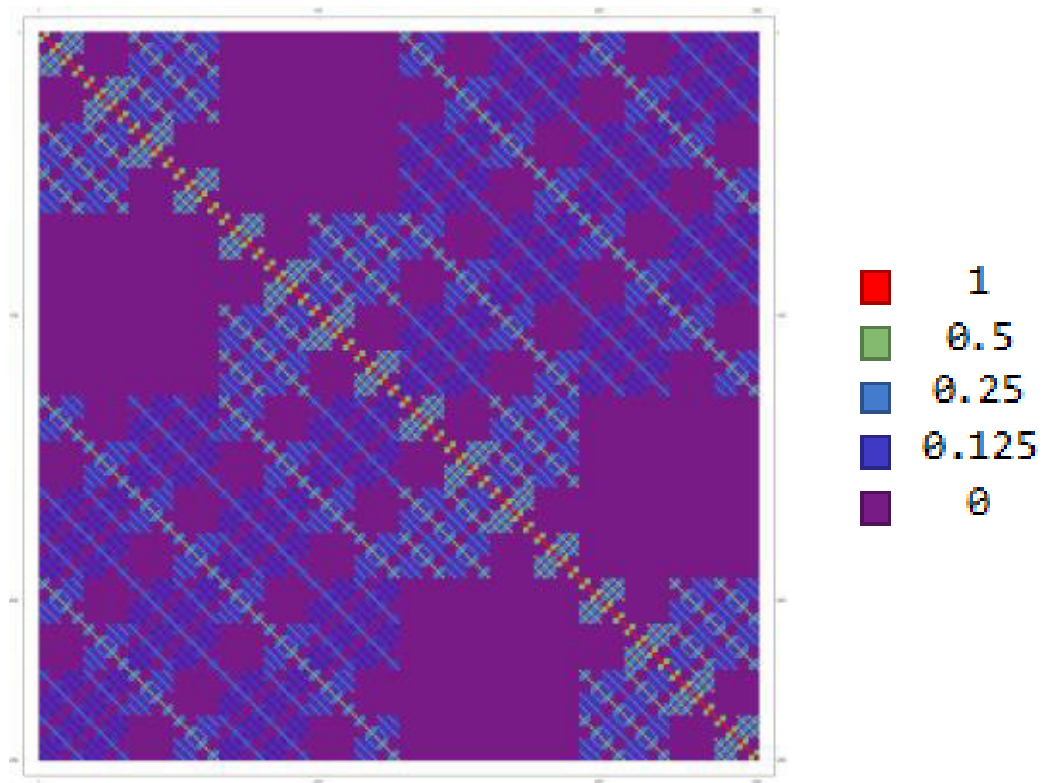

*Fig.11. Probability to find the secret of the secret sharing protocol in the case of four participants. The picture is the same for* $\mathrm{k}_1 = 0$, $\mathrm{k}_1 = \mathrm{k} - 1$ *and* $\mathrm{k}_1 = \lfloor k/2 \rfloor - 1$.

The overall probability of finding a solution for four participants when assuming the initial state is:

$$P_S(\Omega = \Omega_{MAX}, \mathrm{k}_1 = 0) = \frac{1}{16}\sum_M \frac{P_M(\Omega_{MAX}) N_M(\Omega_{MAX})}{256^2} = \frac{1}{16} \tag{119}$$

### 6.5. Semiempirical formula in case of an arbitrary number of participants

Dan has prepared the secret message by encoding it according to $(117)$ with $k_1 - 1$ iterations. Let there be an eavesdropper Eve (E) that intercepts all qubits.

Let Eve has access to the “padlock”. To do the decryption, she needs to "guess" the correct initial state. The probability to randomly guess it is $4^{-Q}$, where Q is the number of participants. All real and assumed initial states of each qubit can be expressed as $|S_l\rangle_2 = \left(|0\rangle + e^{i\varphi_l}|1\rangle\right)/\sqrt{2}$ and $|S'_l\rangle_2 = \left(|0\rangle + e^{i\varphi'_l}|1\rangle\right)/\sqrt{2}$. Each angle $\varphi'_l$ and $\varphi_l$ has four possible values $\varphi'_l, \varphi_l \in \{0, \pi/2, \pi, 3\pi/2\}$. There are only four options for their difference:

$$\begin{gathered} \Delta\varphi_l = \varphi'_l - \varphi_l = 0 \\ \Delta\varphi_l = \varphi'_l - \varphi_l = \pi/2 \\ \Delta\varphi_l = \varphi'_l - \varphi_l = \pi \\ \Delta\varphi_l = \varphi'_l - \varphi_l = 3\pi/2 \end{gathered} \qquad (120)$$

Let us denote the number of mistaken phases for which $\Delta\varphi_l = \pi/2$ and $\Delta\varphi_l = \pi$ respectively as follows:

$$\Xi[\pi/2] = \sum_{l=0}^{Q} \left(\delta_{\Delta\varphi_l,\pi/2} + \delta_{\Delta\varphi_l,3\pi/2}\right) \qquad (121)$$

$$\Xi[\pi] = \sum_{l=0}^{Q} \delta_{\Delta\varphi_l,\pi} \qquad (122)$$

Based on our analytical simulations for two and three participants, as well as numerical simulations with up to and including seven participants, we estimate that the probability of obtaining the secret depending on the number of errors is:

$$P(S_j, S'_j) = \left(\frac{1}{2}\right)^{\Xi[\pi/2]} \frac{1 + Sign\left[1.5 - 2^{\Xi[\pi]}\right]}{2} \qquad (123)$$

*Table.5* shows how the probability $P_M$ of obtaining the secret $\Omega = \Omega_{MAX}$ depends on the phases mistaken by the eavesdropper. The ratio of all element combinations $N_M$ that that have probability $P_M$ is also shown. Each error with phase π/2 decreases twice the probability to obtain the secret. If there are one or more errors with phase π, this reduces the probability of obtaining the secret to zero.

| № | Condition | $P_M(\Omega = \Omega_{MAX})$ | $N_M(\Omega = \Omega_{MAX})$ |
|---|---|---|---|
| 0 | $\left\|\varphi'_j - \varphi_j\right\| = 0 \forall j \in \{1,2,3, \dots Q\}$ | 1 | $\frac{1}{4^Q}$ |
| 1 | $\left\|\varphi'_j - \varphi_j\right\| = \frac{\pi}{2}$<br>$\left\|\varphi'_k - \varphi_k\right\| = 0 \forall\, k \neq j$ | 0.5 | $\frac{Q2^Q}{4^Q} = \frac{Q}{2^Q}$ |
| 2 | $\left\|\varphi'_{j1} - \varphi_{j1}\right\| = \left\|\varphi'_{j2} - \varphi_{j2}\right\| = \frac{\pi}{2}$<br>$\left\|\varphi'_k - \varphi_k\right\| = 0 \forall\, k \neq j_1, j_2$ | 0.25 | $\frac{Q(Q-1)2^{Q-1}}{4^Q} = \frac{Q(Q-1)}{2^{Q+1}}$ |
| 3 | $\left\|\varphi'_{jl} - \varphi_{jl}\right\| = \pi/2 \forall\, l = 1,2,3$<br>$\&\left\|\varphi'_k - \varphi_k\right\| = 0 \forall\, k \neq j_1, j_2, j_3$ | 0.125 | $\frac{Q(Q-1)(Q-2)2^{Q-2}}{4^Q}$<br>$= \frac{Q(Q-1)(Q-2)}{2^{Q+2}}$ |
| | … | | |
| | $\left\|\varphi'_{jl} - \varphi_{jl}\right\| = \pi/2 \forall\, l = 1,2, \dots, r$<br>$\&\left\|\varphi'_k - \varphi_k\right\| = 0 \forall\, k \neq j_1, \dots, j_r$ | $2^{-r}$ | $\frac{Q!}{(Q-r)!\, 2^{Q+r}}$ |
| | … | | |
| Q+1 | $\exists\, j \text{ such that } \left\|\varphi'_j - \varphi_j\right\| = \pi$ | 0 | $-\sum_{i=1}^{Q} 4^{Q-i} \prod_{j=1}^{i} \frac{j-Q-1}{i!}$<br>$= 1 - \left(\frac{3}{4}\right)^Q$ |

*Table.5.Probabilitis for an eavesdropper to obtain the secret depending on the mistaken phases. The second column shows the required conditions to obtain the probability $P_M$ (shown in the third column). The fourth column shows the proportion of cases where these conditions are met.*

In summary, the probability for Eve to obtain the secret via an interception attack is the same as by random guessing.

$$P_M(\Omega = \Omega_{MAX}) = \frac{1}{4^Q} \sum_M P_M(\Omega_{MAX}) N_M(\Omega_{MAX}) = \frac{1}{2^Q} \qquad (124)$$

The eavesdropper may decide to complete only the current iteration and then do a measurement, but even in this case she needs to guess the initial state. The probability of obtaining the secret in this case is the same as by random guessing.

## 7. Possible areas were this protocol is applicable

Like Grover's algorithm, the corresponding secret sharing protocol consists of a sequence of generalized Householder reflections. Such reflections can easily be implemented in various physical systems, like photonic or ion trap-based quantum computers.

The main disadvantage of the Grover cryptographic protocol in the case when the qubits are not distributed among the participants during the last Grover iteration, is the need of padlock device that must be used to complete Grover's algorithm. This protocol can only be used if there is a place where the key can be stored securely. Examples include bank vaults, corporate organizations, and other.

Grover's cryptographic protocol, when qubits are distributed in the final iteration of Grover, does not require such a padlock. It can be used in the same way as other secret-sharing protocols, for example for secure distributed data storage.

It is also important to note that here we check the security of the protocol only against interception attack. We cannot guarantee that this modification is secure against other types of attacks.

## 8. Conclusion

In this work, we introduce a secret sharing protocol we introduce a secret sharing protocol based on a modification of the Grover's search algorithm and study its security against interception attacks. Analytical formulas for the probability of an eavesdropper to obtain the secret via an interception attack are obtained are obtained for an arbitrary value of the phase of the generalized Householder reflections used in Grover's search algorithm, including the case with zero failure rate of the algorithm. We also construct a generalization of this protocol for any number of participants and analyze its security against interception attacks. In particular we deduce an analytical formula for the probability for an eavesdropper to obtain the secret when the distribution of the qubits between the participants during the first and during the last Grover iteration respectively. For the cases when the distribution is during the first or last iteration, we show analytically that this probability is the same as the probability of guessing the secret at random, which proves that in those cases the protocol is secure against interception attacks. Numerical simulations were done for the cases of four, five, six and seven participants. Based on the analytical and numerical results, we construct a generalization of this protocol for any number of participants and distribution during arbitrary Grover iteration and extrapolate to obtain semiempirical formulas for the security of the protocol. Finally, we briefly compared the advantages and limitations of qubit distribution during the first or during the last iteration.